\documentclass[a4paper]{article}
\pdfoutput=1
\usepackage{amsmath,amsfonts,amsthm,amssymb}
\usepackage{mathtools}
\usepackage{bm}
\usepackage{booktabs} 
\usepackage[usenames,dvipsnames]{xcolor}
\usepackage{tikz-cd}
\usetikzlibrary{matrix,arrows,calc}
\usepackage{calc}
\usepackage[colorlinks=true,linkcolor=blue,citecolor=red,urlcolor=Violet,bookmarksdepth=subsection,final]{hyperref}
\usepackage[margin=3cm]{geometry}
\usepackage{authblk}

\makeatletter
\renewcommand*\env@matrix[1][*\c@MaxMatrixCols c]{%
  \hskip -\arraycolsep
  \let\@ifnextchar\new@ifnextchar
  \array{#1}}
\makeatother

\tikzset{ampersand replacement=\&}

\allowdisplaybreaks[1] 

\newtheorem{thm}{Theorem}

\newtheorem{prp}[thm]{Proposition}

\theoremstyle{remark}

\def \bk {\mbox{\boldmath{$k$}}}
\def \bv {\mbox{\boldmath{$v$}}}
\def \bl {\mbox{\boldmath{$\ell$}}}
\def \bn {\mbox{\boldmath{$n$}}}
\def \bm {\mbox{\boldmath{$m$}}}
\def \bA {\mbox{\boldmath{$A$}}}
\def \bF {\mbox{\boldmath{$F$}}}
\def \bT {\mbox{\boldmath{$T$}}}

\newcommand{\mathi}{\mathrm{i}}
\newcommand{\mathe}{\mathrm{e}}

\newcommand{\eqend}[1]{\,#1}
\newcommand{\bigo}[1]{\mathcal{O}\!\left({#1}\right)}

  \newcommand{\del}{\partial}
  \newcommand{\oo}{\infty}
  
\newcommand{\total}{\mathop{}\!\mathrm{d}}
\renewcommand{\d}{\total}

\renewcommand{\r}{\mathsf{r}}
  \newcommand{\R}{\mathsf{R}}
  \newcommand{\rH}{\r_\text{H}}
  \newcommand{\N}{\mathsf{N}}
  \newcommand{\m}{\mathsf{m}}
  
  \newcommand{\om}{\varpi}

  \def \bF {\mbox{\boldmath{$F$}}}

  \def \bk {\mbox{\boldmath{$k$}}}
  \def \bl {\mbox{\boldmath{$\ell$}}}

  \usepackage{alphalph}

  \def\@spliteq#1{\begin{equation}\begin{split}#1\end{split}\end{equation}}
  \def\splitequation{\collect@body\@spliteq}
  
  \makeatother

\title{On well-posedness and algebraic type of the five-dimensional charged rotating black hole with two equal-magnitude angular momenta}

\author[1]{Markus B. Fr\"ob}
\author[2,3]{Igor Khavkine}
\author[2]{Tom\'a\v{s} M\'alek}
\author[2]{Vojt\v{e}ch Pravda}

\affil[1]{Institut f\"ur Theoretische Physik, Universit\"at Leipzig,
	Br\"uderstra{\ss}e 16, 04103 Leipzig, Germany}
\affil[2]{Institute of Mathematics of the Czech Academy of Sciences, \v{Z}itn\'a 25,
	115 67 Prague 1, Czechia}
\affil[3]{Charles University, Faculty of Mathematics and Physics, Sokolovsk\'a 83, 186 75 Prague 8, Czechia}
\date{}

\begin{document}
\maketitle

\vspace*{-3mm}
E-mail: \href{mailto:mfroeb@itp.uni-leipzig.de}{mfroeb@itp.uni-leipzig.de},
\href{mailto:khavkine@math.cas.cz}{khavkine@math.cas.cz},
\href{mailto:malek@math.cas.cz}{malek@math.cas.cz}, \\
\hspace*{17mm} \href{mailto:pravda@math.cas.cz}{pravda@math.cas.cz} \\[2mm]


\begin{abstract}
	 We study various mathematical aspects of the charged rotating black hole with two equal-magnitude angular momenta in five dimensions. We introduce a coordinate system that is regular on the horizon and in which Einstein--Maxwell equations reduce to an autonomous system of ODEs. Employing Bondi and Kruskal-like coordinates, we analyze the geometric regularity of the black hole metric at infinity and the horizon, respectively, and the well-posedness of the corresponding boundary value problem.

	 We also study the algebraic types of the electromagnetic and curvature tensors. While outside the horizon the electromagnetic and Ricci tensors are of type D, the Weyl tensor is algebraically general. The Weyl tensor simplifies to type~II on the horizon and type~D on the bifurcation sphere. These results imply inconsistency of the metric with the Kerr--Schild form with a geodesic Kerr--Schild vector. This feature is shared by the four-dimensional Kerr--Newman metric and the vacuum Myers--Perry or charged Schwarzschild--Tangherlini geometries in arbitrary dimension, but hence not by the black hole we have considered here.
\end{abstract}

\section{Introduction} \label{sec:intro}

Black holes represent the most basic objects of general relativity in four and higher dimensions. It is relatively straightforward to find static, spherically symmetric black hole solutions to vacuum Einstein or Einstein--Maxwell equations. However, it is much more difficult to generalize such black hole solutions to the rotating case and usually, an additional simplifying assumption about the metric has to be employed on top of stationarity and axial symmetry.

In particular, in constructing the Kerr metric in \cite{Kerr63}, the metric was assumed to be of Petrov type D.
For the construction of a vacuum higher-dimensional rotating black hole metric \cite{MyePer86}, the essential assumption was that it can be cast in the Kerr--Schild form, similarly as in the four-dimensional case:
\begin{equation}
g_{\mu\nu} = \eta_{\mu\nu} - 2 \mathcal{H} k_{\mu} k_{\nu} \label{GKS} \eqend{,}
\end{equation}
where $\eta_{\mu \nu}$ is the Minkowski metric, $\mathcal{H}$ a scalar function and $\bk$
is a null vector with respect to $\eta_{\mu \nu}$~\footnote{It follows that $\bk$ is also null with respect to the full metric $g_{\mu \nu}$.}.

The above two assumptions---Petrov type D and the Kerr--Schild form---in fact also hold for the four-dimensional charged rotating black hole described by the Kerr--Newman metric \cite{NewmanJanis65,NewmanCouch65} and for charged Schwarzschild--Tangherlini black holes \cite{Tangherlini63} in arbitrary dimension.
Furthermore, in these cases, the null eigenvectors of the electromagnetic field $\bF$ coincide with the principal null directions of the Weyl tensor \cite{OrtPra06}. Note also that for Kerr, Kerr--Newman, Myers--Perry, and Schwarzschild--Tangherlini metrics, the Kerr--Schild vector $\bk$ is geodesic.

In contrast with the four-dimensional case, where the Kerr--Newman solution was constructed in less than a year after the publication of the Kerr metric, the exact solution representing a higher-dimensional charged rotating black hole is now, three and a half decades after the publication of the Myers--Perry solution \cite{MyePer86}, still unknown\footnote{Note that various approximate solutions, for example for small charge \cite{AliFro04,nl-5d}, slow rotation \cite{Aliev06,Aliev06prd} or large dimensions \cite{Andrade19}, are known.}.
Being a 5-dimensional object, such a black hole is not of direct
relevance to the emerging field of gravitational wave astrophysics, which
has accompanied the development of LIGO, LISA and other gravitational
wave observatories. However, it may be of interest to more detailed
future phenomenological investigations of new physics scenarios like
brane world models, where 5-dimensional gravity interacts with matter
confined to a 4-dimensional brane~\cite{gregory-bwbh}. Independent of
their phenomenological significance, charged rotating black holes are a
good theoretical testing ground for methods of interpolating between
known charged static and uncharged rotating solutions, which besides
higher dimensions may become useful also in modified theories of gravity
in 4~dimensions.

In this paper, we focus on the charged rotating black hole in five dimensions with two equal-magnitude angular
momenta, starting with the metric ansatz previously employed in the numerical studies of this black hole, initiated in~\cite{Kunz2005,kunz-etal} and continued in~\cite{nl-5d, fan-liang-mei}.
These works have demonstrated the existence (or mathematically speaking,
provided very strong evidence thereof) of a solution to the
Einstein--Maxwell equations with this ansatz, by numerically solving an
ordinary boundary value problem in the region between the horizon and
infinity. Their focus was on studying the numerical relationships
between various black hole parameters that can only be determined from
the knowledge of the global solution, rather than from only local
approximations. For instance, while the horizon radius, surface gravity,
rotation speed, and surface electric potential at the horizon are all
quantities local to the horizon, and while the total mass, angular
momentum, charge, and gyromagnetic ratio are local to infinity, definite
relations between these two groups of parameters must be globally
determined. However, these previous works have not discussed or have
left implicit certain structural or algebraic properties of these black
holes. It is to these matters that we turn our attention in this work.

Namely, we focus on two main aspects: the geometric regularity of the
black hole solution at infinity and at the horizon, and on the algebraic
type of the metric in the region between these two extremes. Indeed, we
confirm that the black hole metric has a regular extension across the
horizon and that it possesses a regular null infinity without incoming
or outgoing radiation. On the other hand, we also show that in the bulk
the metric is \emph{algebraically general}\footnote{Here we refer to the generalization of the standard Petrov classification of the Weyl tensor in four dimensions to other tensors and to higher dimensions~\cite{Coleyetal04}, \cite{Milsonetal05} (see \cite{OrtPraPra12rev} for a review). \emph{Algebraically general} means that the Weyl tensor does not admit a multiple Weyl aligned null direction (multiple WAND). In contrast, type II and D spacetime admit one and (at least) two multiple WANDs respectively.} and incompatible with
Kerr--Schild form \eqref{GKS} with geodesic $\bk$, so it could not have
appeared in the lists of exact solutions obtained under these
simplifying assumptions. On the other hand, employing Kruskal-like coordinates, we show that
 the metric does become algebraically special at the horizon, in agreement with the geometric horizon conjecture of \cite{Coleyetal17} (cf. also \cite{LewPaw05}). 
 
It is worth noting that while most of our results hold for equal-magnitude angular momenta, the results on the algebraically general Weyl tensor and non-compatibility of the metric with the Kerr--Schild form \eqref{GKS} with geodesic $\bk$ can be obviously extended to the case of generic angular momenta (see \cite{Kunz2005} for the corresponding metric ansatz).

The paper is structured as follows: 
In Section~\ref{sec:ansatz}, we recall the metric ansatz
from~\cite{kunz-etal} and show why it is not optimal for studying the
regularity of the metric at the horizon by matching it to the known
exact uncharged (Myers--Perry) and non-rotating (charged
Schwarzschild--Tangherlini) solutions. In Section~\ref{sec:reg}, we
introduce a different radial coordinate and a modified metric ansatz,
which is better adapted to studying the regularity of the horizon. We
then obtain the corresponding autonomous system of Einstein--Maxwell
equations in the form of a constrained ordinary boundary value problem
with singular end-points. Then a careful comparison of the solution at
the singular end-points with the geometric regularity conditions at
infinity and at the horizon shows that the boundary value problem is indeed
well-posed. This analysis also suggests how to ensure numerical accuracy
in future investigations of these solutions.

In Section~\ref{sec:algebraic}, we then study the simplifying assumptions holding for Kerr, Kerr--Newman, Schwarzschild--Tangherlini, and Myers--Perry metrics and we show that none of these hold for the five-dimensional charged rotating black hole with two equal-magnitude angular momenta.
We show that this black hole is algebraically general. It also turns out that while the Ricci tensor is of type~D and aligned with a type~D Maxwell tensor $\bF$,
the geometric properties of the common null aligned vector imply that the five-dimensional charged rotating black hole is not compatible with the Kerr--Schild ansatz \eqref{GKS} with geodesic $\bk$. In five dimensions the necessary condition for such compatibility boils down to $\hat Q \hat J = 0$ (see Equation \eqref{geodcond2}) which explains why this ansatz holds only for rotating vacuum or charged non-rotating five-dimensional black holes.

We should note that the previous work~\cite{fan-liang-mei} also
attempted to put the charged rotating black hole metric into Kerr--Schild
form~\eqref{GKS}, but they only succeeded by replacing the reference
Minkowski metric $\eta_{ab}$ by a different metric with unclear
algebraic or geometric properties. Moreover, the authors claim that by
using some differential identities, they have simplified the
corresponding Einstein--Maxwell equations to a system on only two unknown
functions, while also giving an asymptotic solution to the resulting
equations. Unfortunately, we have identified a mistake in a key identity
that allowed their simplification and have also found that their
asymptotic solution is not accurate. A more detailed discussion can be
found in Appendix~\ref{sec:flm}.

\section{Metric ansatz} \label{sec:ansatz}

Specializing the metric and vector potential ansatz for equal-magnitude angular momenta black holes in Einstein--Maxwell theory
from~\cite{kunz-etal} to five dimensions, gives
\begin{multline} \label{eq:kunz-ansatz}
\total s^2 = g_{\mu\nu} \total x^\mu \total x^\nu = - f \total t^2 + \frac{m}{f} \left( \total r^2 + r^2 \total \theta^2 \right) \\
	+ \frac{n}{f} r^2 \left[ \sin^2 \theta \left( \total \phi - \frac{\omega}{r} \total t \right)^2 + \cos^2 \theta \left( \total \psi - \frac{\omega}{r} \total t \right)^2 \right] + \frac{m-n}{f} r^2 \sin^2 \theta \cos^2 \theta \left( \total \phi - \total \psi \right)^2
\eqend{,}
\end{multline}
and
\begin{equation}
A = A_\mu \total x^\mu = a_0 \total t + a_\phi \left( \sin^2 \theta \total \phi + \cos^2 \theta \total \psi \right) \eqend{,}
\end{equation}
where the six functions $f$, $m$, $n$, $\omega$, $a_0$ and $a_\phi$ only
depend on the radial coordinate $r$. Ordering the coordinates as
$t,r,\theta,\phi,\psi$, the metric in matrix form is
\begin{equation}
\resizebox{0.93\hsize}{!}{$
g_{\mu\nu} = \frac{1}{f} \begin{pmatrix} - f^2 + n \omega^2 & 0 & 0 & - n \omega r \sin^2 \theta & - n \omega r \cos^2 \theta \\ 0 & m & 0 & 0 & 0 \\ 0 & 0 & m r^2 & 0 & 0 \\ - n \omega r \sin^2 \theta & 0 & 0 & \left( m \cos^2 \theta + n \sin^2 \theta \right) r^2 \sin^2 \theta & (n-m) r^2 \sin^2 \theta \cos^2 \theta \\ - n \omega r \cos^2 \theta & 0 & 0 & (n-m) r^2 \sin^2 \theta \cos^2 \theta & \left( m \sin^2 \theta + n \cos^2 \theta \right) r^2 \cos^2 \theta \end{pmatrix} \eqend{,}
$}
\end{equation}
The outer horizon, located at $r = r_\text{H}$, is determined by the
condition $f(r_\text{H}) = 0$, with the value of $r_\text{H}$ closest to
$r=\oo$. Two sets of boundary conditions are imposed, one at infinity
and one at the horizon. At infinity ($r\to \oo$) we require asymptotic
flatness, namely
\begin{align}
\total s^2 &\sim - \total t^2 + \total r^2 + r^2 \total \theta^2
	+ r^2 \sin^2\theta\, \total\phi^2
	+ r^2 \cos^2\theta\, \total\phi^2 \eqend{,} \\
A &\sim 0 \eqend{,}
\end{align}
where we have followed~\cite{kunz-etal} by not specifying further
details about what higher dimensional asymptotic flatness means (we give
a more detailed definition of asymptotic flatness later in
Section~\ref{sec:af}).
In~\cite{kunz-etal}, the corresponding boundary condition is stated as
the following asymptotic limits:
\begin{align} \label{eq:kunz-bc-inf}
&f, m, n \sim 1 \eqend{,} \\
&\omega, a_0, a_\phi \sim 0 \eqend{.}
\end{align}
At the horizon ($r=r_\text{H}$), we require regularity, which means the
existence of local coordinates in which the metric smoothly extends
through the horizon. In~\cite{kunz-etal}, the corresponding boundary
condition is stated as follows:
\begin{align} \label{eq:kunz-bc-hor}
&f(r_\text{H}), m(r_\text{H}), n(r_\text{H}) = 0 \eqend{,} \\
&|\omega(r_\text{H})|, |a_0|, |a_\phi| < \oo \eqend{,} \\
&a_\phi'(r_\text{H}) = 0 \eqend{.}
\end{align}

Following~\cite[Sec.~3]{kunz-etal},
provided the function $f$, $\omega$ and $a_0$ have asymptotic expansions
at spatial infinity ($r\to \oo$) of the form
\begin{equation} \label{eq:charge-params}
	f = 1 - \frac{\hat{M}}{r^2} + \cdots \eqend{,} \quad
	\omega = \frac{\hat{J}}{r^3} + \cdots \eqend{,} \quad
	a_0 = \frac{\hat{Q}}{r^2} + \cdots \eqend{,}
\end{equation}
the physical mass, angular momentum and electric charges are
respectively given by
\begin{equation} \label{eq:charges}
\begin{aligned}
	M &= \frac{(d-2) A(S^{d-2})}{16 G_d} \hat{M}
		= \frac{3 \pi^2}{8 G_5} \hat{M}
		\eqend{,} \\
	J &= \frac{A(S^{d-2})}{8 G_d} \hat{J}
		= \frac{\pi^2}{4 G_5} \hat{J}
		\eqend{,} \\
	Q &= \frac{(d-3) A(S^{d-2})}{4 G_d} \hat{Q}
		= \frac{\pi^2}{G_5} \hat{Q}
		\eqend{,}
\end{aligned}
\end{equation}
where $d=5$ is the spacetime dimension, $G_d$ is the $d$-dimensional
Newton's constant, and $A(S^{d-2}) = A(S^3) = 2\pi^2$ is the area of the
unit $3$-sphere. The  physical charges $M$ and $J$  are obtained by
integrating the usual Komar forms at spatial infinity.

\subsection{Special exact solutions}
\label{sec:exact_solutions}

As we shall see in Section~\ref{sec:reg}, the radial coordinate implied
by the ansatz~\eqref{eq:kunz-ansatz} has non-differentiable behavior (a
square root singularity) at the horizon $r=r_\text{H}$. This phenomenon
is most efficiently exhibited by matching the ansatz to the known exact
solutions without rotation (5-dimensional charged Schwarzschild--Tangherlini metric) or
without charge (5-dimensional Myers--Perry metric), which we do next. Since we
are interested in making the comparison in the region exterior to the
black hole, we restrict ourselves to the subextremal ranges of the
charge, mass, and angular momentum parameters of the exact
solutions.

Before moving on, let us introduce a new radial coordinate $\r = \r(r)$,
which will play the role of the more familiar radial coordinate in these
exact solutions that regularly extends through the horizon. It so
happens that in all cases considered in this paper, this new radial
coordinate will obey the relationship
\begin{equation} \label{eq:old-r}
	\frac{(\d{r})^2}{r^2} = \frac{\r^2 (\d\r)^2}{(\r^2 - \r_+^2)(\r^2 - \r_-^2)}
	\quad \iff \quad
	r = r_\text{H} \sqrt{\frac{\sqrt{\r^2-\r_-^2} + \sqrt{\r^2-\r_+^2}}{\sqrt{\r^2-\r_-^2} - \sqrt{\r^2-\r_+^2}}} \eqend{,}
\end{equation}
where $\r_\pm$ are some constants and the differential relation has been
solved to match $\r(r_\text{H}) = \r_+$. The square root singularity is
clearly visible at the horizon in the expansion $r = r_\text{H} +
\bigo{\sqrt{\r-\r_+}}$, while at infinity $r = \bigo{\r}$ and has
a regular expansion in integer powers of $\r^{-1}$.

The 5-dimensional \emph{charged Schwarzschild--Tangherlini} (ST) solution
(sometimes also called the 5-dimensional \emph{Reissner--Nordstr\"om} solution)
is given by
\begin{equation}
\begin{gathered}
	\d{s^2} = -f \d{t}^2
		+ \frac{\d\r^2}{f}
		+ \r^2 (\d\theta^2 + \sin^2\theta \d\phi^2 + \cos^2\theta \d\psi^2)
		\eqend{,} \\
	A = \left(\frac{\hat{Q}}{\r^2} - \hat{C}\right) \d{t} \eqend{,} \quad
	f = 1-\frac{\hat{M}}{\r^2} + \frac{4}{3} \frac{\hat{Q}^2}{\r^4}
		= \frac{(\r^2 - \r_+^2) (\r^2 - \r_-^2)}{\r^4} \eqend{,} \\
	\hat{M} = \r_+^2 + \r_-^2 \eqend{,} \quad
	\text{with} \quad
	\hat{Q} = \frac{\sqrt{3}}{2} \r_+ \r_- \eqend{,} \quad
\end{gathered}
\end{equation}
where the mass and charge are parametrized by the constants
$(\hat{M},\hat{Q})$ or $(\r_+, \r_-)$, $\r_+ > \r_-$. Matching the ST
solution with the ansatz~\eqref{eq:kunz-ansatz} gives the radial
coordinate relation~\eqref{eq:old-r} and
\begin{equation}
\begin{aligned}
	f &= \frac{(\r^2 - \r_+^2) (\r^2 - \r_-^2)}{\r^4} \eqend{,} \\
	m &= \frac{\r^2}{r^2} f \eqend{,} \\
	n &= \frac{\r^2}{r^2} f \eqend{,} \\
	\omega &= 0 \eqend{,} \\
	a_0 &= \frac{\sqrt{3}}{2} \frac{\r_+ \r_-}{\r^2} - \hat{C} \eqend{,} \\
	a_\phi &= 0 \eqend{.}
\end{aligned}
\end{equation}

The 5-dimensional \emph{Myers--Perry} (MP) solution with equal-magnitude angular momenta is given by
\begin{equation}
\begin{aligned}
	\d{s^2} &= -\d{t}^2
		+ \frac{\Sigma \r^2 \d\r^2}{\Sigma^2 - \hat{M}\r^2}
		+ \Sigma \r^2 (\d\theta^2 + \sin^2\theta \d\phi^2 + \cos^2\theta \d\psi^2)
			\\ & \qquad {}
		+ \frac{\hat{M}}{\Sigma} \left(\d{t} + a \sin^2\theta \d\phi + a \cos^2\theta \d\psi\right)^2
		\eqend{,} \\
	A &= 0 , \quad
	\Sigma = \r^2 + a^2, \quad
	\Sigma^2 - \hat{M} \r^2 = (\r^2 - \r_+^2) (\r^2 - \r_-^2) \eqend{,} \\
	& \qquad
	\text{with} \quad
	\hat{M} = (\r_+ + \r_-)^2 , \quad
	a = \hat{J}/\hat{M} = \sqrt{\r_+ \r_-} \eqend{,} \quad
\end{aligned}
\end{equation}
where the mass and the angular momentum are parametrized by the
constants $(\hat{M},\hat{J})$ or $(\r_+, \r_-)$, $\r_+ > \r_-$.
Matching the MP solution with the ansatz~\eqref{eq:kunz-ansatz} gives
\begin{equation}
\label{MPmetric}
\begin{aligned}
	f &= \frac{(\r^2 - \r_+^2) (\r^2 - \r_-^2)}
		{(\r^2 + \r_- \r_+)^2 + \r_+ \r_- (\r_- + \r_+)^2}
		= \frac{\Sigma^2 - \hat{M}\r^2}{\Sigma^2 + a^2 \hat{M}}
		\eqend{,} \\
	m &= \frac{(\r^2 + \r_- \r_+)}{r^2} f
		= \Sigma f
		\eqend{,} \\
	n &=
		\frac{(\r^2 - \r_+^2) (\r^2 - \r_-^2)}{r^2 (\r^2 + \r_+\r_-)}
		= \frac{\Sigma^2 - \hat{M}\r^2}{\Sigma}
		\eqend{,} \\
	\omega &= r \frac{\sqrt{\r_+\r_-} (\r_+ + \r_-)^2}
		{(\r^2 + \r_- \r_+)^2 + \r_+ \r_- (\r_- + \r_+)^2}
		= \frac{a \hat{M}}{\Sigma^2 + a^2 \hat{M}} \eqend{,} \\
	a_0 &= 0 \eqend{,} \\
	a_\phi &= 0 \eqend{.}
\end{aligned}
\end{equation}

\section{Regularity at infinity and at the horizon} \label{sec:reg}

Since the radial coordinate $r$ used in the
ansatz~\eqref{eq:kunz-ansatz} has been found to be non-smooth at the
horizons of known exact solutions, we must reparametrize the ansatz
using our new regular coordinate $\r$. In addition, we have found that
removing most of the explicit dependence on $\r$ in the metric makes the
corresponding Einstein equations (given below) autonomous, which is
technically convenient.

Our new \emph{regular} ansatz for the metric and vector potential is
\begin{multline} \label{eq:reg-ansatz}
\total s^2 = g_{\mu\nu} \total x^\mu \total x^\nu =
	- f \total t^2
	+ \frac{\m}{f} \left( \frac{(\r\total \r)^2}{\N} + \total \theta^2 \right) \\
	+ \frac{\N}{\m} \left[\sin^2 \theta \left(\total \phi - \om \total t \right)^2
		+ \cos^2 \theta \left(\total \psi - \om \total t \right)^2 \right]
	+ \frac{\m^2-f\N}{\m f} \sin^2 \theta \cos^2 \theta \left( \total \phi - \total \psi \right)^2
\end{multline}
and
\begin{equation}
A = A_\mu \total x^\mu = a_0 \total t + a_\phi \left( \sin^2 \theta \total \phi + \cos^2 \theta \total \psi \right) \eqend{.} \label{eq:potential_reg}
\end{equation}
The functions $f$, $\m$, $\N$, $a_0$ and $a_\phi$ all depend only on the
new radial coordinate $\r$. In matrix form, ordering the coordinates as
$(t,\r,\theta,\phi,\psi)$, the metric reads
\begin{equation}
g_{\mu\nu} = \begin{pmatrix} - f + \frac{\N}{\m} \om^2 & 0 & 0 & - \frac{\N}{\m} \om \sin^2 \theta & - \frac{\N}{\m} \om \cos^2 \theta \\ 0 & \frac{\r^2\m}{f \N} & 0 & 0 & 0 \\ 0 & 0 & \frac{\m}{f} & 0 & 0 \\ - \frac{\N}{\m} \om \sin^2 \theta & 0 & 0 & \frac{\left( \m^2 \cos^2 \theta + f\N \sin^2 \theta \right)}{f \m} \sin^2 \theta & \frac{(f\N-\m^2)}{f \m} \sin^2 \theta \cos^2 \theta \\ - \frac{\N}{\m} \om \cos^2 \theta & 0 & 0 & \frac{(f\N-\m^2)}{f \m} \sin^2 \theta \cos^2 \theta & \frac{\left( \m^2 \sin^2 \theta + f\N \cos^2 \theta \right)}{f \m} \cos^2 \theta \end{pmatrix} \eqend{.}
\end{equation}
Obviously, compared to~\eqref{eq:kunz-ansatz}, only the metric part has
been reparametrized. We are using a new radial coordinate $\r$ and the
old parametrization is recovered via the translation formulas
\begin{subequations} \label{eq:reg-kunz-ansatz}
\begin{align}
	\m &= m r^2 \eqend{,} \\
	\om &= \frac{\omega}{r} \eqend{,} \\
	\N &= \frac{m n r^4}{f} \eqend{,} \\
	\frac{(\r\total\r)^2}{\N} &= \frac{(\total{r})^2}{r^2} \eqend{.}
\end{align}
\end{subequations}

Note that the radial coordinate $\r$ appears explicitly in the new
ansatz~\eqref{eq:reg-ansatz} only through the combination $\r\total\r =
\frac{1}{2} \total\R$, where $\R = \r^2$. Hence, it is possible and
useful to write the resulting Einstein equations as an autonomous system
of ODEs with respect to the squared radial coordinate $\R$. In
particular, we will see that the way $\N$ appears in the ansatz gives it
a very simple equation of motion. As long as the location of the horizon satisfies
$\r = \rH \ne 0$, if $\r$ is a regular coordinate at the horizon, then
so is $\R$. Again, we are interested in the interval from infinity
$\r\sim \oo$ to the horizon $f(\r=\rH)=0$, the root closest to infinity.

Plugging~\eqref{eq:reg-ansatz} directly into the Einstein equations
gives somewhat complicated expressions. It is unenlightening to present
them directly. Instead, inspired by the simplifications that have
already been noted in~\cite{kunz-etal, nl-5d, fan-liang-mei}, we will
give them in equivalent but more structured form. In what follows, we
use the notation $(-)' = \frac{\d}{\d\R}(-)$ and all functions are
treated as functions of $\R$. First, we have the set of conservation
laws
\begin{subequations} \label{eq:reg-eom-cons}
\begin{align}
	\N'' &= 2
		\implies & \N &= (\R-\r_+^2) (\R-\r_-^2) \eqend{,} \\
	\left(\frac{\N}{f} (a_0' + \om a_\phi')\right)' &= 0
		\implies & \frac{\N}{f} (a_0' + \om a_\phi') &= -\hat{Q} \eqend{,} \\
\label{eq:reg-eom-cons-omega}
	\left(\frac{\N^2}{f \m} \om' + 4 \hat{Q} a_\phi\right)' &= 0
		\implies & \frac{\N^2}{f \m} \om' + 4 \hat{Q} a_\phi &= -2\hat{J} \eqend{,}
\end{align}
\end{subequations}
where $\r_+, \r_-, \hat{Q}, \hat{J}$ have been introduced as integration
constants. Then we have the boundary value problem (BVP)
\begin{subequations} \label{eq:reg-eom-bvp}
\begin{align}
\label{eq:reg-eom-bvp-f}
	f^2 \m \left(\frac{\m}{f^2} f'\right)'
		- \m \m' \left(\frac{\m'}{\m} - 2\frac{\N'}{\N}\right) f
		- \left(4\frac{\m^2}{\N} - f\right) f
	&= \frac{8}{3} \frac{f \m}{\N} \left(2\m^2 (a_\phi')^2 - f a_\phi^2\right)
		\notag \\ &\quad {}
		- \frac{4\hat{Q}^2}{3} \frac{f^2 \m^2}{\N^2} \eqend{,}
		\\
\label{eq:reg-eom-bvp-m}
	f \m \left(\frac{\N}{\m} \m'\right)'
		+ \left(\frac{f^2 \N}{\m^2} - 2f\right) \m
	&= \frac{4}{3} f \left(2\m^2 (a_\phi')^2 - f a_\phi^2\right)
		\notag \\ &\quad {}
		+ \frac{4\hat{Q}^2}{3} \frac{f^2 \m}{\N}
		+ 4 \frac{f^2 \m^2}{\N^2} (\hat{J} + 2\hat{Q} a_\phi)^2 \eqend{,}
		\\
\label{eq:reg-eom-bvp-a}
	\m \left( \m a_\phi' \right)'
		- f a_\phi
	&= 2\hat{Q} \frac{f \m^2}{\N^2} (\hat{J} + 2\hat{Q} a_\phi) \eqend{,}
\end{align}
\end{subequations}
and finally we have the constraint
\begin{multline} \label{eq:reg-eom-constraint}
	\m^2 (a_\phi')^2
	= f a_\phi^2 - \m + \frac{f \N}{4\m} + \hat{Q}^2 \frac{f \m}{\N}
		+ \frac{f \m^2}{\N^2} (\hat{J} + 2\hat{Q} a_\phi)^2
	\\
		+ \frac{\N \m'}{8} \left(\frac{f'}{f}-2\frac{\m'}{\m}+4\frac{\N'}{\N}\right)
		- \frac{\N \m f'}{8 f} \left(2\frac{f'}{f}-\frac{\m'}{\m}+2\frac{\N'}{\N}\right)
	\eqend{.}
\end{multline}
This last constraint is compatible with the BVP system because, letting
$\mathsf{C}$ be the difference between the left- and right-hand sides
of~\eqref{eq:reg-eom-constraint},
\begin{equation} \label{eq:reg-eom-compat}
	(\mathsf{C}/f^2)' = 0 \mod{\text{\eqref{eq:reg-eom-bvp}}} \eqend{,}
\end{equation}
which means that, if the boundary conditions for the BVP system are
chosen such that $\mathsf{C}/f^2 \to 0$ either at infinity or at the
horizon, then the constraint $\mathsf{C}=0$ is guaranteed for any BVP
solution.

Direct substitution shows that when our system of equations is
satisfied, then so are the unsimplified Einstein equations. Conversely,
starting from the unsimplified Einstein equations and eliminating all
variables except $\N$ gives the first conservation law
from~\eqref{eq:reg-eom-cons}. Using it to eliminate $\N''$, as well as
every variable and its derivatives that do not appear in the second
conservation law, gives that same conservation law. Repeating this idea
once more, gives the third conservation law. Upon eliminating $\N$,
$\om$, $a_0$ and their derivatives from the unsimplified equations,
solving for $f'', \m'', a_\phi''$ gives the BVP
system~\eqref{eq:reg-eom-bvp}. Finally eliminating these and all other
second derivatives leaves the constraint
equation~\eqref{eq:reg-eom-constraint}. The derivation of the unsimplified
equations and the above elimination steps are most
easily carried out using computer algebra; we have relied in part on the
xTensor and xCoba packages from~\cite{XACT}.

In the following sections, we will check how the geometric conditions of
asymptotic flatness and regularity at the horizon translate to boundary
conditions for the BVP system~\eqref{eq:reg-eom-bvp}. Under these
conditions, both endpoints become \emph{regular} (or \emph{Fuchsian})
\emph{singular points}, where a solution can be constructed by the
Frobenius method (as a power series in $\R$). At each endpoint, we
perform a Fuchsian analysis (see~\cite{kh-rw} for a brief exposition of
the standard methods from~\cite{wasow}, together with references to the
larger literature), which counts the free parameters in the space of
solutions compatible with the boundary condition. This analysis will
help us confirm that the Einstein equations restricted to our ansatz are
indeed well-posed. Well-posedness is of course a prerequisite property
for trying to construct either numerical or asymptotic approximations to
high accuracy. Unfortunately, the original papers~\cite{kunz-etal,
nl-5d} did not report such a well-posedness analysis, nor specify in
what way their numerical methods were compatible with the expected
asymptotic structure of the solution at the horizon and infinity, which
in principle may have led to spurious numerical artifacts near those
points.

\subsection{Asymptotic flatness} \label{sec:af}

There are different notions of asymptotic flatness both in four and higher
dimensions. We will refer specifically to the existence of a regular
null infinity, where both asymptotic Poincar\'e symmetries and their
charges are well defined. Specifically in five~dimensions, this notion was
analyzed in~\cite{tts-5d} and translated to precise component
asymptotics for the metric in Bondi coordinates, which is what we will
refer to below. Later, the same analysis was generalized to arbitrary
higher dimension in~\cite{tts-higherd}. Earlier work treated asymptotic
flatness at null infinity in even dimensions~\cite{hi-evend} and at
spatial infinity~\cite{tts-i0} using conformal methods. A unified
treatment of the higher dimensional asymptotics  of scalar,
electromagnetic and gravitational fields at null infinity was given
in~\cite{sw-asymp}, which is the main reference that we will follow.
Along similar lines, \cite{ht-asymp-maxwell} treated
higher dimensional asymptotics for electromagnetic fields at both null
and spatial infinities. Independently, various asymptotic behaviors for
electromagnetic fields in higher dimensions were also analyzed
in~\cite{ortaggio-em}.

Effectively summarizing the results of the above references (most
clearly presented in~\cite[Sec.~II~B]{sw-asymp} for electromagnetic and
in~\cite[Sec.~II~D--E]{sw-asymp} for gravitational fields), the
conditions of having an asymptotically flat regular (past/future) null
infinity, with the absence of (incoming/outgoing) radiation (both
gravitational and electromagnetic), but with non-vanishing contributions
to total mass and electric charge, is equivalent to the existence of an
asymptotic coordinate system $(u,r,\theta,\phi,\psi)$ and corresponding
electromagnetic gauge where
\begin{subequations} \label{eq:bondi-af}
\begin{multline}
	g_{\mu\nu} =
	\begin{pmatrix}
		-1 & \pm 1 & 0 & 0 & 0 \\
		\pm 1 & 0 & 0 & 0 & 0 \\
		0 & 0 & r^2 & 0 & 0 \\
		0 & 0 & 0 & r^2\sin^2\theta & 0 \\
		0 & 0 & 0 & 0 & r^2\cos^2\theta
	\end{pmatrix} \\
	+
	\begin{pmatrix}
		\bigo{r^{-2}}
			& \bigo{r^{-2}}
			& \bigo{r^{-1}}
			& \bigo{r^{-1}}
			& \bigo{r^{-1}} \\
		\bigo{r^{-2}}
			& 0 & 0 & 0 & 0 \\
		\bigo{r^{-1}}
			& 0 & \bigo{r^0} & \bigo{r^0} & \bigo{r^0} \\
		\bigo{r^{-1}}
			& 0 & \bigo{r^0} & \bigo{r^0} & \bigo{r^0} \\
		\bigo{r^{-1}}
			& 0 & \bigo{r^0} & \bigo{r^0} & \bigo{r^0}
	\end{pmatrix}
\end{multline}
and
\begin{equation}
	A_\mu = \begin{pmatrix}
		\bigo{r^{-2}} & 0 & \bigo{r^{-1}} & \bigo{r^{-1}} & \bigo{r^{-1}}
	\end{pmatrix} \eqend{.}
\end{equation}
\end{subequations}
Some explanation is in order. The $\pm$ choice of sign in the metric
distinguishes the past and future null infinities. The exact zero
entries in $g_{\mu\nu}$ and $A_\mu$ are part of the gauge fixing
conditions. The Bondi gauge~\cite{tts-5d} usually also requires that the
angular submatrix $g_{IJ}$, where $I,J$ run through $\theta, \phi,
\psi$, has determinant exactly equal to $r^6 \sin^2\theta \cos^2\theta$
(where $\sin^2\theta \cos^2\theta$ is the determinant of the unit round
$3$-sphere metric in the same coordinates). However, it is also
sufficient that this subdeterminant has this limit only for $r \to \infty$, with subleading
terms given by an asymptotic expansion in integer powers of $r$, because
then the $r$ coordinate can be redefined to obtain exact equality,
without affecting the structure of the asymptotics. Below, when adapting
our ansatz to the Bondi gauge, we will not perform this extra
redefinition, in favor of our more convenient $\r$ coordinate. However,
such a redefinition would of course be possible. Note also that the
Bondi $r$ coordinate is logically distinct from the coordinate denoted
by the same symbol in the metric~\eqref{eq:kunz-ansatz}. Since the Bondi
radial coordinate is only used in this section, there should be no
confusion between the two.

In dimension $d=5$, in the absence of gravitational radiation, the
leading radial \emph{Coulombic} asymptotic for both gravitational and
electromagnetic perturbations is $\bigo{r^{-d+3}} = \bigo{r^{-2}}$ in an
orthonormal basis~\cite{sw-asymp}%
	\footnote{Reference~\cite{sw-asymp} uses harmonic instead of Bondi
	gauge. But from their discussion, it is clear that one can transform
	from one gauge to the other without changing the leading asymptotics.}%
, which explains different decay rates in different coordinate tensor
components above. Coulombic terms are those that can contribute to
finite and non-vanishing mass, angular momentum and electric charges. In
the presence of gravitational and electromagnetic radiation, the
perturbations would decay only as $\bigo{r^{-d/2+1}} = \bigo{r^{-3/2}}$
in an orthonormal basis~\cite{sw-asymp}. It should also be mentioned
that the leading asymptotic terms will be further constrained by the
Einstein equations themselves, which may lead to faster decay for some
components, as we shall see in our case.

We can define the desired Bondi (which could also be called Eddington--Finkelstein) coordinates as
\begin{equation} \label{eq:ef-coords}
\begin{aligned}
	\d{u_\pm} &= \d{t} \pm \sqrt{\frac{\m}{\N}} \frac{\d\R}{2 f} \eqend{,} \\
	r &= \r \eqend{,} \\
	\theta &= \theta \eqend{,} \\
	\d\phi_\pm &= \d\phi \pm \om \sqrt{\frac{\m}{\N}} \frac{\d\R}{2 f} \eqend{,} \\
	\d\psi_\pm &= \d\psi \pm \om \sqrt{\frac{\m}{\N}} \frac{\d\R}{2 f} \eqend{.}
\end{aligned}
\end{equation}
In the coordinates $(u_\pm,\r,\theta,\phi_\pm,\psi_\pm)$,
the matrix form of the metric is
\begin{subequations} \label{eq:ef-tensors}
\begin{equation} \label{eq:ef-metric}
g_{\mu\nu} = \begin{pmatrix} - f + \frac{\N}{\m} \om^2 & \pm\sqrt{\frac{\r^2\m}{\N}} & 0 & - \frac{\N}{\m} \om \sin^2 \theta & - \frac{\N}{\m} \om \cos^2 \theta \\ \pm\sqrt{\frac{\r^2\m}{\N}} & 0 & 0 & 0 & 0 \\ 0 & 0 & \frac{\m}{f} & 0 & 0 \\ - \frac{\N}{\m} \om \sin^2 \theta & 0 & 0 & \frac{\left( \m^2 \cos^2 \theta + f\N \sin^2 \theta \right)}{f\m} \sin^2 \theta & \frac{(f\N-\m^2)}{f\m} \sin^2 \theta \cos^2 \theta \\ - \frac{\N}{\m} \om \cos^2 \theta & 0 & 0 & \frac{(f\N-\m^2)}{f\m} \sin^2 \theta \cos^2 \theta & \frac{\left( \m^2 \sin^2 \theta + f\N \cos^2 \theta \right)}{f\m} \cos^2 \theta \end{pmatrix} \eqend{,}
\end{equation}
while the vector potential takes on the form
\begin{equation} \label{eq:ef-A}
	A_\mu - (\d \alpha)_\mu
		= \begin{pmatrix}
			a_0 & 0 & 0 & a_\phi\sin^2\theta & a_\phi\cos^2\theta
		\end{pmatrix}
		\eqend{,}
\end{equation}
\end{subequations}
with the gauge transformation parameter defined by $\alpha'(\r) =
-\sqrt{\r^2\m/\N} (a_0 + \om a_\phi)/f$.

Hence, comparing~\eqref{eq:ef-tensors} with~\eqref{eq:bondi-af}, both
the metric and the vector potential are in desired Bondi form, provided
we require the following leading asymptotics (recalling that $\R=\r^2$)
\begin{equation} \label{eq:af-limits}
\begin{aligned}
	f &= 1 + \bigo{\R^{-1}} \eqend{,} \\
	\m &= \R + \bigo{\R^0} \eqend{,} \\
	a_\phi &= \bigo{\R^{-1/2}} \eqend{,} \\
	\N &= \R^2 + \bigo{\R^1} \eqend{,} \\
	\om &= \bigo{\R^{-3/2}} \eqend{,} \\
	a_0 &= \bigo{\R^{-1}} \eqend{.}
\end{aligned}
\end{equation}
Note that comparing with~\eqref{eq:bondi-af} has allowed us to determine
 not only the leading but also the subleading terms of $f$,
$\m$ and $\N$. We have not yet specified the structure of the full
asymptotic expansion in~\eqref{eq:af-limits}, since in principle we can
allow subleading terms with arbitrary fractional powers of $\R$.
However, we are also free to restrict the subleading terms as we see
fit, with the only justification necessary the a~posteriori
well-posedness of the BVP part of the Einstein equations,
\eqref{eq:reg-eom-bvp} and~\eqref{eq:reg-eom-cons}.

So we may assume that the asymptotic expansion is in integer powers of
$\R$, which is consistent because we are solving equations that are
autonomous with respect to $\R$, and perform a Fuchsian analysis of the
BVP system at the singular point $\R=\oo$. For linear equations, this is
the same as applying the method of Frobenius. For a non-linear equation,
the first step is take the leading terms from~\eqref{eq:af-limits} and
check that they are consistent with the BVP
system~\eqref{eq:reg-eom-bvp}. Namely, the leading order terms are
collected and it is checked that they cancel among themselves. We have
specifically written~\eqref{eq:reg-eom-bvp} in a way that all leading
terms appear on the left-hand side. The cancellation, which is verified
by direct calculation, could have imposed some conditions on the
undetermined leading coefficient in $a_\phi$, but it turns out that it
does not.

The second step is to linearize~\eqref{eq:reg-eom-bvp} and apply the
method of Frobenius to the linear equation whose coefficients have
asymptotics determined by the leading terms~\eqref{eq:af-limits}. If
$\R=\oo$ turns out to be a regular singular point of this linear
equation, we can extract the corresponding indicial equation and
determine the remaining free coefficients in the expansion. For systems
of ODEs, determining whether a singular point is regular is not entirely
trivial. The procedure must take into account that the leading terms in
different components of the system of unknowns or of the system of
equations may be of different orders, let alone that the components of
the equations and of the unknowns are allowed to mix.

There is a convenient criterion to check when a system of $k$ equations
of order $p$ has a regular singular point (again, see~\cite{kh-rw} for a
brief exposition). Let $E[v] = 0$ be such a system of linear equations,
whose coefficients have asymptotic expansions in powers of $\R$. Let $S$
and $T$ be $k\times k$ matrices such that all of $S$, $T$, $S^{-1}$,
$T^{-1}$ have components that are Laurent polynomials in $\R$ (for that,
it is necessary and sufficient that $\det S$ and $\det T$ are
non-vanishing monomials). The system has a regular singular point at
$\R=\oo$ if $T^{-1}E[S v] = E_0(\R\,\del_\R)[v] + \text{lower order terms}$, where the
components of $E_0(x)$ are constant coefficient polynomials in $x$,
$\det E_0(x) \ne 0$, and for the purposes of collecting lower order
terms $\R^n (\R\,\del_\R)^m$ has order $n$ (which will typically be
negative). We call matrices $S$ and $T$ \emph{leading multipliers}. The
integer solutions of the \emph{indicial equation} $\det E_0(n) = 0$ are
the critical exponents (or indices) of $\R^n$ whose coefficients in the
expansion of the solution may be a free parameter. The eigenvectors of
$E_0(n)$ corresponding to the critical exponents determine how these
free parameters enter the expansion. At a true real regular singular
point, the polynomial degree of $\det E_0(n)$ should be $pk$, determined
by the size and order of the system. If the degree is lower, then only a
subset of the solutions exhibit a Frobenius expansion in powers of $\R$
(with possible logarithmic contributions) and there will exist other
solutions that grow or decay faster than simple powers of $\R$.

The third and last step counts the number of free parameters in the
asymptotic expansion of a solution. If $v = S \left[\sum_{k>n} v_k \R^n
+ v_n \R^n + O(\R)^{n-1}\right]$, then knowing that the $v_{k>n}$
coefficients solve the ODE system to appropriate order we can solve for
the next coefficient, provided that $E_0(n)$ is invertible, via
\begin{equation}
	E_0(n) v_n = e_n(v_{k>n}) \eqend{,}
\end{equation}
where $e_n(-)$ collects all the remaining terms of the original non-linear
ODE system at order $n$. The invertibility of $E_n(n)$ fails
precisely when $n$ is a critical exponent. Then two things happen:
because $E_0(n)$ does not have full row rank, the consistency of the
equation imposes constraints on $e_n(v_{k>n})$ and hence on any free
parameters appearing among the $v_{k>n}$ coefficients; also, because
$E_0(n)$ does not have full column rank, new free parameters appear in
the solution for the $v_n$ coefficients. In general the number of free
parameters gained and lost at a time need not be the same, so at each
critical exponent the total number of free parameters in the expansion
may increase, decrease, or stay the same; it stabilizes after the last
critical exponent.

In general, leading multiplier matrices may be quite complicated, but we
have specifically written the BVP system~\eqref{eq:reg-eom-bvp} to make
them simple. Namely, the resulting $E_0(\R\,\del_\R)[v]$ operator is
\begin{equation}
	\begin{bmatrix}
		(\R\,\del_\R)^2 + 1 & 2(\R\,\del_\R - 1) & 0 \\
		1 & (\R\,\del_\R + 2) (\R\,\del_\R - 1) & 0 \\
		0 & 0 & (\R\,\del_\R + 1) (\R\,\del_\R - 1)
	\end{bmatrix}
	\begin{bmatrix}
		\delta f \\
		\frac{\delta \m}{\R} \\
		\delta a_\phi
	\end{bmatrix} \eqend{,}
\end{equation}
where $S$ multiplies $\delta \m$ by $\R^{-1}$ and $T^{-1}$ multiplies
the middle equation by $\R$, while acting as the identity on other
components. The critical exponents are $k = 1~(2), 0~(1), -1~(3)$, with
multiplicities indicated in parentheses. Because of the constraints on
the leading terms from~\eqref{eq:af-limits}, the coefficients
corresponding to the $k=1,0$ exponents must vanish. At $k=-1$, one free
parameter is allowed in front of the leading term of $a_\phi$ and two
more free parameters appear in the subleading terms of $f$ and $\m$,
with algebraic multiplicity $2$. As is usual in applying the Frobenius
method, the algebraic multiplicity $2$ implies that the two independent
coefficients appear in front of $\R^{-1}$ and $\R^{-1} \log\R$ in the
expansions of $f$ and $\m/\R$. However, the logarithmic terms are
incompatible with the asymptotics~\eqref{eq:af-limits} corresponding to
a regular null infinity. Thus, only one free coefficient at order
$\R^{-1}$ remains. The remaining two free parameters remain free at all
further orders of the expansion.

Once the asymptotic expansions for $f,\m,a_\phi$ have been determined,
we can also asymptotically integrate the first order conservation
laws~\eqref{eq:reg-eom-cons} for $\om$ and $a_0$. For each of them,
there is a free parameter that appears as an additive constant.

In summary, the following leading terms uniquely fix
the most general asymptotic expansion of a solution in powers of $\R$
that is compatible with the asymptotic form~\eqref{eq:af-limits}:
\begin{equation} \label{eq:reg-sol-inf}
\begin{aligned}
	f  &= 1 + \frac{f^{(-1)}}{\R} + \bigo{\R^{-2}} \eqend{,} \\
	\m &= \R + \m^{(0)} + \bigo{\R^{-1}} \eqend{,} \\
	a_\phi &= \frac{a_\phi^{(-1)}}{\R} + \bigo{\R^{-2}} \eqend{,} \\
	\N &= \R^2 - (\r_+^2+\r_-^2) \R + \r_+^2 \r_-^2 \eqend{,} \\
	\om &= \om^{(0)} + \frac{\hat{J}}{\R^2} + \bigo{\R^{-3}} \eqend{,} \\
	a_0 &= a_0^{(0)} + \frac{\hat{Q}}{\R} + \bigo{\R^{-2}} \eqend{,}
\end{aligned}
\end{equation}
which has exactly two free parameters ($a_\phi^{(-1)}$ and a linear
combination of $f^{(-1)}$ and $\m^{(0)}$), since the yet undetermined
coefficients are constrained by
\begin{equation} \label{eq:reg-sol-inf-compat}
\begin{aligned}
	f^{(-1)} - 2 \m^{(0)} - (\r_+^2+\r_-^2) &= 0 \eqend{,} \\
	\om^{(0)} &= 0 \eqend{,} \\
	a_0^{(0)} &= 0 \eqend{,}
\end{aligned}
\end{equation}
after enforcing compatibility with the prescribed asymptotic limits of
$f, \m, \om$ and $a_0$. Comparing with~\eqref{eq:charge-params}
justifies the notation $\hat{J}$ and $\hat{Q}$ for the coefficients in
the expansion of $\omega$ and $a_0$ and also allows us to identify
\begin{equation} \label{eq:M-from-f}
	f^{(-1)} = -\hat{M} \eqend{,}
\end{equation}
with $\hat{M}$, $\hat{J}$ and $\hat{Q}$ being proportional to the
physical mass, angular momentum and electric charges, respectively,
as specified in~\eqref{eq:charges}.
Finally, there are no further constraints coming from enforcing the
limit $\mathsf{C}/f^2 \to 0$, which is needed to ensure that the
constraint $\mathsf{C}=0$~\eqref{eq:reg-eom-cons} is enforced by the
compatibility equation~\eqref{eq:reg-eom-compat}. Recall that $\r_+^2,
\r_-^2$ are integration constants that appear
in~\eqref{eq:reg-eom-cons}. Note that, after we have carefully taken into
account the Einstein--Maxwell equations, the final leading asymptotics $a_\phi =
\bigo{\R^{-1}}$ and $\om = \bigo{\R^{-2}}$ decay faster than strictly
required by asymptotic flatness in~\eqref{eq:af-limits}.

Having shown that at infinity the series expansion for the solution is
uniquely fixed as a function of $\r_\pm$, $\hat{J}$, $\hat{Q}$,
$a_\phi^{(-1)}$ and say $\hat{M}$, we report here the first few terms
of $f$, $\m$, and $a_\phi$ extending~\eqref{eq:reg-sol-inf} (defining
$\rH^2 = \r_+^2 - \r_-^2)$:
\begin{equation} \label{eq:reg-sol-inf4}
\begin{aligned}
	f &= 1 - \frac{\hat{M}}{(\R-\r_-^2)}
		+ \frac{\frac{1}{2} \hat{M} \left(\hat{M}-\rH^2\right)+\frac{4}{3} \hat{Q}^2}{(\R-\r_-^2)^2}
		\\ & \quad {}
		+ \frac{\frac{4}{9} \left( a_\phi^{(-1)} \right)^2-\frac{1}{6} \hat{M} \left(\hat{M}^2-3 \hat{M} \rH^2+2 \rH^4\right)+\frac{4}{9} \hat{Q}^2 \left(-4\hat{M}+3 \rH^2\right)+\frac{2}{3} \hat{J}^2}{(\R-\r_-^2)^3}
		\\ & \quad {}
		+ \frac{\frac{2}{9} \left( a_\phi^{(-1)} \right)^2 \left(-2 \hat{M}+3 \rH^2\right)+\frac{16}{9} a_\phi^{(-1)} \hat{J} \hat{Q}+\frac{1}{9} \hat{Q}^2 \left(11 (\hat{M}-\rH^2)^2-2\hat{M} \rH^2\right)}{(\R-\r_-^2)^4}
		\\ & \qquad {}
		+ \frac{\frac{1}{24} \hat{M} (\hat{M} - \rH^2) (\hat{M} - 2\rH^2) (\hat{M} - 3\rH^2)+\hat{J}^2 \left(-\frac{7}{6} \hat{M}+\rH^2\right)+\frac{32}{27} \hat{Q}^4}{(\R-\r_-^2)^4}
		\\ & \quad {}
		+ \bigo{\R^{-5}} \eqend{,} \\
	\m &= (\R-\r_-^2)
		- \frac{1}{2} \left(\hat{M}+\rH^2\right)
		+ \frac{\frac{4}{3} \hat{Q}^2}{(\R-\r_-^2)}
		\\ & \quad {}
		+ \frac{\frac{1}{72} \left(16 \left( a_\phi^{(-1)} \right)^2+3 \hat{M}^3-3 \hat{M} \rH^4\right)+\frac{2}{9} \hat{Q}^2 \left(-4 \hat{M}+3 \rH^2\right)+\frac{5}{6} \hat{J}^2}{(\R-\r_-^2)^2}
		\\ & \quad {}
		+ \frac{\frac{1}{360} \left(-\hat{M}+5 \rH^2\right) \left(16 \left( a_\phi^{(-1)} \right)^2+3 \hat{M}^3-3 \hat{M} \rH^4\right)+\frac{88}{45} a_\phi^{(-1)} \hat{J} \hat{Q}}{(\R-\r_-^2)^3}
		\\ & \qquad {}
		+ \frac{\frac{1}{45} \hat{Q}^2 \left(9 \hat{M}^2-40 \hat{M} \rH^2+23 \rH^4\right)+\hat{J}^2 \left(-\frac{29}{30} \hat{M}+\frac{5}{6} \rH^2\right)+\frac{128}{135} \hat{Q}^4}{(\R-\r_-^2)^3}
		\\ & \quad {}
		+ \bigo{\R^{-4}} \eqend{,} \\
	a_\phi &= \frac{a_\phi^{(-1)}}{(\R-\r_-^2)}
		+ \frac{\frac{2}{3} \hat{J} \hat{Q}+\frac{1}{6} a_\phi^{(-1)} \left(\hat{M}+3 \rH^2\right)}{(\R-\r_-^2)^2}
		\\ & \quad {}
		+ \frac{\frac{1}{12} a_\phi^{(-1)} \left(\hat{M}^2+2 \hat{M} \rH^2+3 \rH^4\right)+\frac{1}{6} \hat{J} \hat{Q} \left(-\hat{M}+4 \rH^2\right)}{(\R-\r_-^2)^3}
		\\ & \quad {}
		+ \frac{-\frac{1}{360} a_\phi^{(-1)} \left(16 \left( a_\phi^{(-1)} \right)^2-6 \hat{M}^3-45 \hat{M}^2 \rH^2-48 \hat{M} \rH^4-45 \rH^6\right)}{(\R-\r_-^2)^4}
		\\ & \qquad {}
		+ \frac{-\frac{34}{135} a_\phi^{(-1)} \hat{M} \hat{Q}^2-\frac{7}{30} a_\phi^{(-1)} \hat{J}^2+\frac{1}{60} \hat{J} \hat{Q} \left(5 \hat{M}^2-15 \hat{M} \rH^2+34 \rH^4\right)+\frac{8}{135} \hat{J} \hat{Q}^3}{(\R-\r_-^2)^4}
		\\ & \quad {}
		+ \bigo{\R^{-5}} \eqend{.}
\end{aligned}
\end{equation}
As discussed in more detail in Section~\ref{sec:well-posed}, our reduced
Einstein--Maxwell equations possess the shift symmetry $\R \mapsto \R +
\R_0$, which also shifts the constants $\r_\pm^2 \to \r_\pm^2 + \R_0$.
The above expansion was obtained under the simplifying assumption
$\r_-^2=0$ and then converted to a form that is manifestly invariant
under this shift symmetry. Elsewhere in the text, we will use $\rH$ to
denote the location of the outer horizon in general. So, strictly
speaking, the numerical value of $\rH$ in the above expansion should
only be interpreted as the location of the horizon under the assumption
$\r_-^2=0$.

This expansion is used several times in Section~\ref{sec:algebraic} to
test whether some differential constraints forced by special algebraic
types would be satisfied on-shell, that is, satisfying the
Einstein--Maxwell equations and our boundary conditions. For instance,
deducing the condition~\eqref{geodcond2} required an expansion at least to the order
given above.

\subsection{Regularity at the horizon} \label{sec:horizon}

The vanishing $f(\rH)$ on the horizon signals a coordinate singularity
there, as is expected for coordinates adapted to a timelike Killing
vector when it becomes null. The regularity of the
metric~\eqref{eq:kunz-ansatz} and the corresponding vector potential at
the horizon must then be checked in other coordinates that penetrate the
horizon, where the only requirement for a tensor field to be regular is
for it to have smooth components in that coordinate system.

Traditionally, the simplest horizon penetrating coordinates on a black
hole are of Eddington--Finkelstein type. The
coordinates~\eqref{eq:ef-coords} that we have introduced to play the
role of the Bondi frame at null infinity are actually also of
Eddington--Finkelstein type at the horizon. The $(u_+, \R, \theta,
\phi_+,\psi_+)$ coordinates regular at the future horizon and $(u_-, \R,
\theta, \phi_-,\psi_-)$ ones regular at the past horizon. We will show
below the horizon regularity of our improved
ansatz~\eqref{eq:reg-ansatz} with respect to these coordinates.
Unfortunately, neither of these coordinate systems is regular at the
bifurcation sphere where the two horizons intersect. Thus we will also
introduce Kruskal-like coordinates that cover a neighborhood of the
bifurcation sphere and show that our improved ansatz is regular there as
well.

To reproduce the horizon behavior~\eqref{eq:kunz-bc-hor} imposed in the
original work~\cite{kunz-etal}, we must set $f = \m = \N = 0$ at
$\R=\rH^2$. As a consequence of the integration of the equation of
motion~\eqref{eq:reg-eom-cons} for $\N$, we must set $\rH = \r_+^2$,
where we have required $\r_+\ne 0$ and ordered the integration constants
as $\r_+^2 > \r_-^2$. Combining these requirements with regularity at
the horizon, via smoothness of the tensor
components~\eqref{eq:ef-tensors} in Eddington--Finkelstein coordinates,
gives the horizon asymptotics
\begin{subequations} \label{eq:reg-bc-hor}
\begin{align}
	f &= \bigo{ \R-\rH^2 } \eqend{,} \\
	\m &= \bigo{ \R-\rH^2 } \eqend{,} \\
	a_\phi &= \bigo{ 1 } \eqend{,} \\
	\N &= (\r_+^2-\r_-^2)(\R-\rH^2) + \bigo{ (\R-\rH^2)^2 } \eqend{,} \\
	\om &= \bigo{ 1 } \eqend{,} \\
	a_0 &= \bigo{ 1 } \eqend{,}
\end{align}
\end{subequations}
with each function also smooth in $\R$ at the horizon and where the
leading coefficients of $f$ and $\m$ are non-vanishing. Again, the
asymptotics of $\N$, $\om$ and $a_0$ are obtained by integrating the
conservation laws~\eqref{eq:reg-eom-cons}, which remain consistent with
the regularity of the tensor coefficients in~\eqref{eq:ef-tensors}.

Kruskal (or Kruskal--Szekeres) coordinates for Schwarzschild spacetime
are well-known. It  is not as easy to find analogous coordinates
constructed for the Kerr and related black holes. The original
construction by Carter~\cite{carter-coords} was restricted to the
$(t,r)$ plane along the rotation axis. Pretorius and
Israel~\cite{pretorius-israel} seem to have been the first to construct
global double-null coordinates regular on the Kerr bifurcation sphere.
Motivated by applications in the global non-linear stability of Kerr and
related black holes, analogous double-null coordinates have been
constructed also for a wider class of geometries,
see~\cite{franzen-girao, arganaraz-moreschi} and references therein.

For our purposes, since we are interested only in a neighborhood of the
bifurcation sphere, it is easiest to follow the idea from the much
simpler construction by Hayward~\cite{hayward}. Namely, consider a
constant $\kappa$, which we will choose appropriately, and the
coordinates $(U,V,\theta, \Phi, \Psi)$ defined by
\begin{equation} \label{eq:kruskal-coords}
\begin{aligned}
	\frac{\d{U}}{U} - \frac{\d{V}}{V} &= \kappa \d{t} \eqend{,}
		& \d{t} &= \frac{1}{\kappa}
				\left(\frac{\d{U}}{U} - \frac{\d{V}}{V}\right) \eqend{,} \\
	\frac{\d{U}}{U} + \frac{\d{V}}{V} &= \kappa \sqrt{\frac{\m}{\N}} \frac{\d\R}{2f} \eqend{,}
		& \sqrt{\frac{\m}{\N}} \frac{\d\R}{2f} &= \frac{1}{\kappa}
				\left(\frac{\d{U}}{U} + \frac{\d{V}}{V}\right) \eqend{,} \\
	\d\Phi &= \d\phi - \om(\rH^2) \d{t} \eqend{,}
	& \d\phi &= \d\Phi + \frac{\om(\rH^2)}{\kappa} \left(\frac{\d{U}}{U} - \frac{\d{V}}{V}\right) \eqend{,} \\
	\d\Psi &= \d\psi - \om(\rH^2) \d{t} \eqend{,}
	& \d\psi &= \d\Psi + \frac{\om(\rH^2)}{\kappa} \left(\frac{\d{U}}{U} - \frac{\d{V}}{V}\right) \eqend{.}
\end{aligned}
\end{equation}
Integrating the second equation, we find
\begin{multline}
	\d\ln(UV)
	= \frac{\kappa}{2f'(\rH^2)} \sqrt{\frac{\m'(\rH^2)}{\N'(\rH^2)}}
		\frac{\d\R}{\R-\rH^2}
		+ \bigo{ (\R-\rH^2)^0 }
	\\
	\iff \quad
	UV = (\R/\rH^2 - 1) h(\R) \eqend{,}
	\quad h(\R) = 1 + \bigo{ \R-\rH^2 } \eqend{,}
\end{multline}
provided we choose the overall integration constant appropriately and set
\begin{equation}
	\kappa = 2 f'(\rH^2) \sqrt{\frac{\N'(\rH^2)}{\m'(\rH^2)}} \eqend{.}
\end{equation}

In these new Kruskal coordinates, denoting $\Delta \om = \om -
\om(\rH^2)$, the metric takes the form
\begin{equation}
\label{eq:reg-ansatz-kruskal}
\resizebox{0.93\hsize}{!}{$
g_{\mu\nu} = \begin{pmatrix}
	V^2 \frac{(\Delta\om)^2 \N}{\kappa^2 U^2 V^2 \m} &
		\frac{2f}{\kappa^2 UV} - UV \frac{(\Delta\om)^2 \N}{\kappa^2 U^2 V^2 \m} & 0 &
		-V \frac{(\Delta\om) \N}{\kappa UV \m} \sin^2\theta &
		-V \frac{(\Delta\om) \N}{\kappa UV \m} \cos^2\theta \\
	\frac{2f}{\kappa^2 UV} - UV \frac{(\Delta\om)^2 \N}{\kappa^2 U^2 V^2 \m} &
		U^2 \frac{(\Delta\om)^2 \N}{\kappa^2 U^2 V^2 \m} & 0 &
		U \frac{(\Delta\om) \N}{\kappa UV \m} \sin^2\theta &
		U \frac{(\Delta\om) \N}{\kappa UV \m} \cos^2\theta \\
	0 & 0 & \frac{\m}{f} & 0 & 0 \\
	-V \frac{(\Delta\om) \N}{\kappa UV \m} \sin^2\theta &
		U \frac{(\Delta\om) \N}{\kappa UV \m} \sin^2\theta & 0 &
		\frac{\left( \m^2 \cos^2 \theta + f\N \sin^2 \theta \right)}{f\m} \sin^2 \theta &
		\frac{(f\N-\m^2)}{f\m} \sin^2 \theta \cos^2 \theta \\
	-V \frac{(\Delta\om) \N}{\kappa UV \m} \cos^2\theta &
		U \frac{(\Delta\om) \N}{\kappa UV \m} \cos^2\theta & 0 &
		\frac{(f\N-\m^2)}{f\m} \sin^2 \theta \cos^2 \theta &
		\frac{\left( \m^2 \sin^2 \theta + f\N \cos^2 \theta \right)}{f\m} \cos^2 \theta \end{pmatrix} \eqend{.}
$}
\end{equation}
Recalling that $\Delta \om, f, \m, \N = \bigo{ \R-\rH^2 } = \bigo{UV}$, with the
leading terms of $f$ and $\m$ non-vanishing, all the components and also
\begin{equation}
\det g = -\frac{4 \m \N}{\kappa^4 U^2 V^2} \cos^2\theta \to -\frac{4
\rH^4 \m'(\rH^2) \N'(\rH^2)}{\kappa^4} \cos^2\theta
\end{equation}
remain finite as $\R \to \rH^2$. Hence, under the conditions already imposed
by~\eqref{eq:reg-bc-hor}, the metric is regular in the Kruskal
coordinates $(U,V,\theta,\Phi,\Psi)$ in a neighborhood of the
bifurcation sphere $U=V=0$, where the future and past horizons
$\R=\rH^2$ intersect. The vector potential transforms to
\begin{multline}
	A = \frac{\Delta(a_0 + \om a_\phi)}{\kappa UV}
			\left(V\d{U} - U\d{V}\right)
		+ a_\phi (\sin^2\theta \d\Phi + \cos^2\theta \d\Psi) \\
      + \total \left[ \frac{a_0(\rH^2) + \om(\rH^2) a_\phi(\rH^2)}{\kappa} \ln \frac{U}{V} \right] \eqend{,}
\end{multline}
where we have used the shortcut notation
\begin{equation}
\Delta(a_0 + \om a_\phi) = \left(a_0 + \om(\rH^2) a_\phi\right) - \left(a_0(\rH^2) + \om(\rH^2) a_\phi(\rH^2)\right) \eqend{.}
\end{equation}
As discussed in the case of Eddington--Finkelstein-like coordinates, the
last term of the vector potential is singular but pure gauge (it is exact), while the rest
of the terms are manifestly regular.

Explicitly solving the coordinate transformation, we find
\begin{equation}
\begin{aligned}
	U &= \sqrt{\R/\rH^2 - 1} \,\mathe^{\kappa t/2} \sqrt{h} \eqend{,} &
		\d{U} &= \frac{\kappa}{2} \sqrt{\R/\rH^2 - 1} \,\mathe^{\kappa t/2} \sqrt{h}
			\left(\sqrt{\frac{\m}{\N}}\frac{\d\R}{2f} + \d{t}\right) \eqend{,} \\
	V &= \sqrt{\R/\rH^2 - 1} \,\mathe^{-\kappa t/2} \sqrt{h} \eqend{,} &
		\d{V} &= \frac{\kappa}{2} \sqrt{\R/\rH^2 - 1} \,\mathe^{-\kappa t/2} \sqrt{h}
			\left(\sqrt{\frac{\m}{\N}}\frac{\d\R}{2f} - \d{t}\right) \eqend{,} \\
	\Phi &= \phi - \om(\rH^2) t \eqend{,} &
		\d\Phi &= \d\phi - \om(\rH^2) \d{t} \\
	\Psi &= \psi - \om(\rH^2) t \eqend{,} &
		\d\Psi &= \d\psi - \om(\rH^2) \d{t} \eqend{.} \\
\end{aligned}
\end{equation}

Next, we repeat the Fuchsian analysis of $\R=\rH^2$ as a regular
singular point of the BVP part~\eqref{eq:reg-eom-bvp} of the Einstein--Maxwell
equations restricted to our ansatz, to determine the remaining free
parameters in the Taylor expansion of the solution. Since the steps of
the procedure were already explained in Section~\ref{sec:af}, we merely
summarize the results. For convenience, we will denote $\rho \equiv \R-\rH^2$.

Plugging the leading terms~\eqref{eq:reg-bc-hor} into the BVP
system~\eqref{eq:reg-eom-bvp}, a direct calculation shows that they are
compatible, without any new restrictions on the undetermined
coefficients. Again, we have written the BVP system in such a way that
all cancelling leading terms appear on the left-hand side. Upon
linearization, we find that the leading multiplier matrices $S$ and $T$
may just be taken to be identity. The resulting $E_0(\rho\del_\rho)[v]$
operator is
\begin{equation}
	\begin{bmatrix}
		(\rho \del_\rho-3)(\rho \del_\rho-1) &
			\frac{f^{(1)}}{\m^{(1)}} (\rho\del_\rho - 1) & 0 \\
		0 & (\rho\del_\rho - 1)^2 & 0 \\
		-\frac{a_\phi^{(0)}}{(\m^{(1)})^2}
			- \frac{2\hat{Q} (\hat{J} + 2\hat{Q} a_\phi^{(0)})}{(\r_+^2-\r_-^2)^2} &
			- \frac{4f^{(1)}\hat{Q} (\hat{J} + 2\hat{Q} a_\phi^{(0)})}{\m^{(1)} (\r_+^2-\r_-^2)^2} &
			(\rho \del_\rho)^2
	\end{bmatrix}
	\begin{bmatrix}
		\delta f \\
		\delta \m \\
		\delta a_\phi
	\end{bmatrix} \eqend{.}
\end{equation}
Exchanging the first two rows and columns, we see that the matrix is
lower triangular, so that we can immediately read off its eigenvalues and
determinant (the indicial equation). The critical $\rho^k$ exponents are
$k=0~(2), 1~(3), 3~(1)$, with multiplicities indicated in parentheses.
Note that regularity at the horizon excludes any logarithmic terms like
$\rho^k (\log\rho)^l$ in the expansion. Multiplicities in the critical
exponents that generate the same eigen-vectors do not generate more free
parameters. At $k=0$ we already have the free leading coefficient of
$a_\phi$. At $k=1$, we have the free leading coefficients of $f$ and
$\m$. At $k=3$ we have one new free coefficient in the expansion of $f$
and no restrictions on the other free coefficients. Finally, enforcing
the limit $\mathsf{C}/f^2 \to 0$ needed to ensure that the constraint
$\mathsf{C}=0$~\eqref{eq:reg-eom-constraint} is enforced by the
compatibility equation~\eqref{eq:reg-eom-compat} does not bring in any
new restrictions on the free parameters. Once the Taylor series for
$f,\m,a_\phi$ have been determined, we can also integrate the first
order conservation laws~\eqref{eq:reg-eom-cons} for $\om$ and $a_0$. For
each of them, there is a free parameter that appears as an additive
constant. In summary, the following leading terms uniquely fix the
Taylor series of a solution in powers of $\rho = \R-\rH^2$, with all
free parameters explicitly shown:
\begin{equation} \label{eq:reg-sol-hor}
\begin{aligned}
	f  &= f^{(1)} \rho
		+ \cdots + f^{(3)} \rho^3 + \bigo{\rho^4} \eqend{,} \\
	\m &= \m^{(1)} \rho + \bigo{\rho^2} \eqend{,} \\
	a_\phi &= a_\phi^{(0)} + \bigo{\rho} \eqend{,} \\
	\N &= (\r_+^2 - \r_-^2) \rho + \rho^2 \eqend{,} \\
	\om &= \om^{(0)} + \bigo{\rho} \eqend{,} \\
	a_0 &= a_0^{(0)} + \bigo{\rho} \eqend{.}
\end{aligned}
\end{equation}
The additive constants $\om^{(0)}$ and $a_0^{(0)}$ are not fixed by the
requirements of regularity at the horizon~\eqref{eq:reg-bc-hor}.
Instead, they must be fixed by enforcing compatibility with the
prescribed limits~\eqref{eq:af-limits} at infinity, as was discussed at
the end of Section~\ref{sec:af}. Thus, the parameters $\om^{(0)}$ and
$a_0^{(0)}$ are uniquely fixed, but their values cannot be obtained
locally at the horizon.

\subsection{Well-posedness} \label{sec:well-posed}

In this section we discuss the well-posedness of the BVP
system~\eqref{eq:reg-eom-bvp} with the boundary conditions prescribed by
asymptotic flatness (Section~\ref{sec:af}) and regularity at the horizon
(Section~\ref{sec:horizon}), as well as the overall uniqueness of the
corresponding charged rotating black hole solution.

The BVP system~\eqref{eq:reg-eom-bvp} is a second order ODE system for
three unknowns, with external parameters $\r_+^2$, $\r_-^2$, $\hat{J}$
and $\hat{Q}$. With the external parameters fixed, its solution space is
hence 6-dimensional (counting the free initial data at any regular
point, for instance). The conclusion of Section~\ref{sec:af} is that
there is a 2-dimensional space of solutions compatible with asymptotic
flatness (parametrized by the constants $f^{(-1)}, \m^{(0)},
a_\phi^{(-1)}$~\eqref{eq:reg-sol-inf}, with one
constraint~\eqref{eq:reg-sol-inf-compat} among them). The conclusion of
Section~\ref{sec:horizon} is that there is a 4-dimensional space of
solutions compatible with regularity at the horizon, parametrized by the
constants $f^{(1)}, f^{(3)}, \m^{(1)}, a_\phi^{(0)}$ in Equation~\eqref{eq:reg-sol-hor}.
Hence, within the overall 6-dimensional solution space, the intersection
of these 2- and 4-dimensional families will generically be
0-dimensional, meaning that the solution if it exists is expected to be
locally unique (under small perturbations of free parameters at either
boundary). In other words, the boundary conditions prescribed by
Sections~\ref{sec:af} and Section~\ref{sec:horizon} define a well-posed
BVP for the non-linear system~\eqref{eq:reg-eom-bvp}.

The BVP system~\eqref{eq:reg-eom-bvp} itself depends on 4 external
parameters, the integration constants $\r_+^2, \r_-^2, \hat{J}, \hat{Q}$
from the conservation laws~\eqref{eq:reg-eom-cons}. Two of them,
$\hat{J}$ and $\hat{Q}$, have direct physical interpretations
as being proportional to the total angular momentum
and total charge, respectively~\eqref{eq:charges}. Because the total Einstein--Maxwell
system of equations restricted to the ansatz~\eqref{eq:reg-ansatz} is
autonomous with respect to $\R$, the coordinate shift $\R \to \R + \R_0$
is a symmetry, since it is also compatible with both the horizon and
asymptotic boundary conditions. However, being a simple coordinate
transformation it does not change the isometry class of the solution.
Note that this shift affects the integration constants as $\r_\pm^2 \to
\r_\pm^2 + \R_0$. We can use the shift symmetry to place the horizon at
any convenient location, for example by setting $\r_-^2=0$, which we
will assume from now on. This choice places the horizon at $\rH = \r_+$.
Having decided the location of the horizon, by the well-posedness of the
BVP discussed in the previous paragraph, the solution is (locally)
uniquely fixed by the three remaining parameters. In particular, the
total mass can be written as $\hat{M} = \hat{M}(\rH^2, \hat{J},
\hat{Q})$, where we can use the relation~\eqref{eq:M-from-f} between
$f^{(-1)}$ and $\hat{M}$. Physically, it is more convenient to
dynamically invert this relationship to $\rH^2 = \rH^2(\hat{M}, \hat{J},
\hat{Q})$.

In summary, we see that the isometry class of a charged rotating black
hole with ansatz~\eqref{eq:reg-ansatz} and satisfying the
Einstein--Maxwell equations is (locally) uniquely determined by its total
mass $\hat{M}$, total angular momentum $\hat{J}$, and total charge
$\hat{Q}$.

\begin{figure}
\begin{center}
	\includegraphics[width=.8\textwidth]{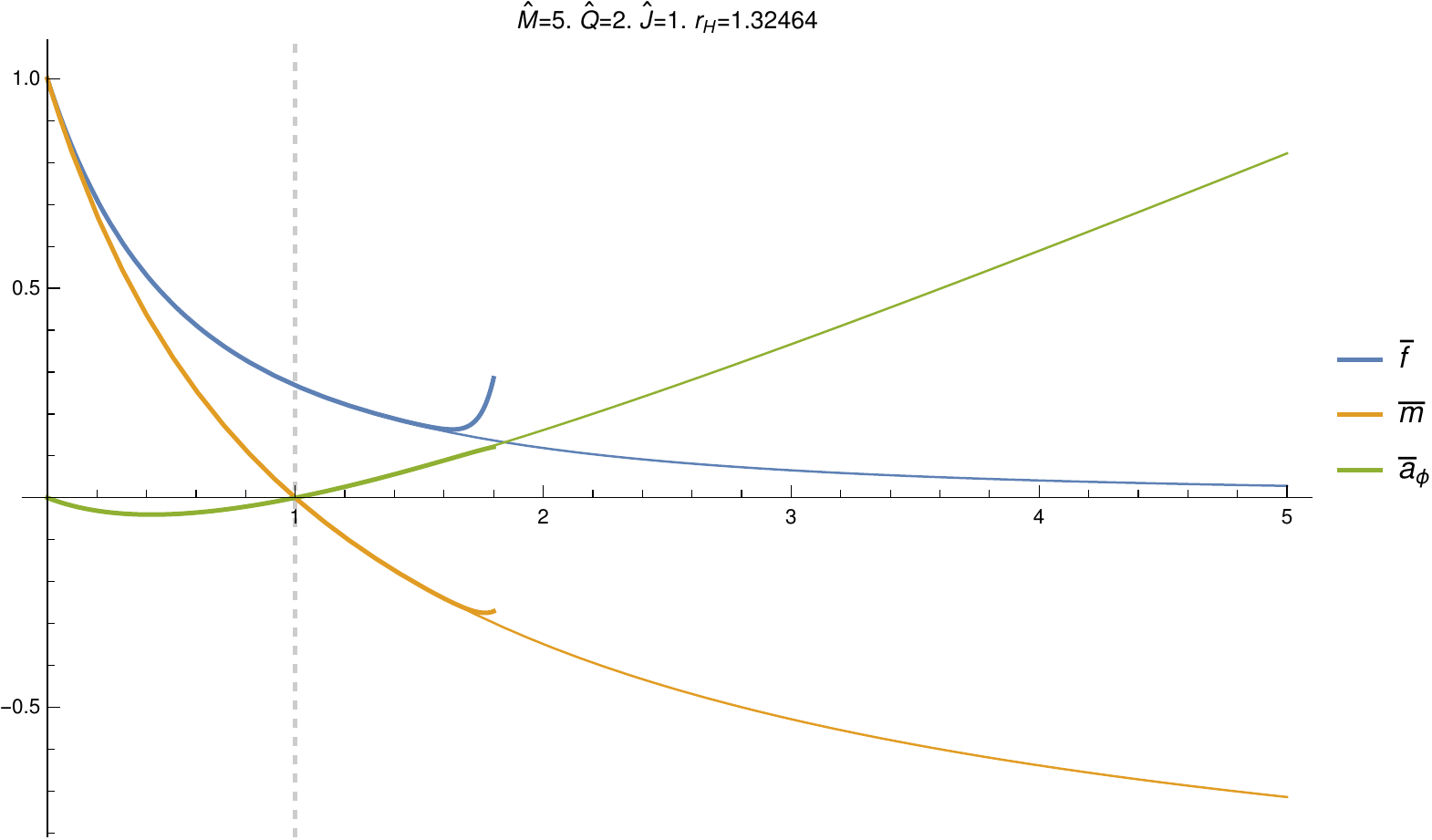}
\end{center}
\caption{Numerical solution for the black hole with external parameters
$\hat{M}=5$, $\hat{J}=1$, $\hat{Q}=2$. The horizon radius is dynamically
determined to be $\rH \approx 1.32464$. The horizontal variable is
$\bar{\R} = \rH^2/\R$, with $\bar{\R}=0$ corresponding to infinity and
$\bar{\R} = 1$ to the horizon, indicated by the dashed vertical line.
The barred functions have been rescaled as $\bar{f} = f/(1-\bar{R})$,
$\bar{\m} = \m \bar{\R}/\rH^2$ and $\bar{a}_\phi = a (1-\bar{\R})$. The
thick curves show the exterior solution and its extrapolation some
distance inside the horizon. The thin curves show the interior
solution.}
\label{fig:num-sol}
\end{figure}

As proof of concept, taking the above well-posedness discussion into
account, we have implemented a numerical solver for the BVP
system~\eqref{eq:reg-eom-bvp} that takes the external $\hat{M}$,
$\hat{J}$ and $\hat{Q}$ parameters as input and constructs the
corresponding solution of the Einstein--Maxwell equations both outside
and inside the horizon. On the exterior region, we use BVP solver in a
basis of Chebyshev polynomials that are weighted at infinity and the
horizon by the expected asymptotics and boundary conditions. Once the
solution on the exterior region has converged, its values and
derivatives near the horizon are used to feed into a standard ODE solver
for the interior region. Figure~\ref{fig:num-sol} illustrates the
resulting solution for generic values of the external parameters. In
particular, we see that even a naive extrapolation of the exterior
solution inside the horizon shows rather good agreement with the
interior solution on a sizable neighborhood of the horizon.

\section{Algebraic classification} \label{sec:algebraic}

In this section, we study algebraic types of  curvature tensors and of the electromagnetic tensor in the charged Myers--Perry metric.

Recall that four-dimensional Kerr and Kerr--Newman black holes are of Petrov type D and both metrics admit a Kerr--Schild form \eqref{GKS} with geodesic $\bk$. Furthermore, for the Kerr--Newman black hole, the null eigenvectors of the electromagnetic tensor $\bF$ coincide with the principal null directions of the Weyl tensor \cite{OrtPra06} (\emph{i.e.}, Weyl tensor and $\bF$ are aligned).

Although the vacuum Myers--Perry black hole is of type D \cite{Pravdaetal04,Hamamotoetal06} and admits Kerr--Schild form with geodesic $\bk$, it has been pointed out already in \cite{MyePer86}, that its charged generalization  does not admit the Kerr--Schild form assuming that the vector potential $\bA$ coincides (up to normalization) with $\bk$. As a consequence of the algebraically special form of the Ricci tensor in the charged Myers--Perry black hole, we will also arrive at the incompatibility of the charged Myers--Perry black hole metric with Kerr--Schild ansatz   \eqref{GKS} with geodesic $\bk$, but without employing the strong constraint $\bA\propto \bk$.

For determining the compatibility of the charged Myers--Perry metric with the Kerr--Schild form with geodesic $\bk$, we will employ the following proposition from \cite{OrtPraPra09}:

\begin{prp}[Geodesicity of the Kerr--Schild vector $\bk$ \cite{OrtPraPra09}]
	\label{prop_GKS_geod}
	The null vector $\bk$ in the Kerr--Schild  metric~\eqref{GKS} is geodesic if, and only if, it is an aligned null direction of the Ricci tensor, \emph{i.e.}, if $R_{ab} k^a k^b = 0$.
\end{prp}
Note (see \cite{OrtPraPra09}) that if $\bk$ in the Kerr--Schild metric~\eqref{GKS} is geodesic, then $\bk$ is also a multiple aligned Weyl direction (WAND) of the spacetime, and thus the spacetime is algebraically special. Since we also show in this section that the charged Myers--Perry metric is not algebraically special, this could be used as another argument against compatibility with the Kerr--Schild form with geodesic $\bk$.

We also discuss the algebraic properties of the test electromagnetic field in the Myers--Perry background and compare this field with the weak field limit of the charged Myers--Perry metric.

\subsection{Algebraic type of Maxwell and Ricci tensors}

The Maxwell tensor obtained from the potential~\eqref{eq:potential_reg} reads
\begin{equation}
F_{\mu\nu} =  \begin{pmatrix} 0 & -a_0' & 0 & 0 & 0 \\ a_0' & 0 & 0 & a_\phi' \sin^2 \theta & a_\phi' \cos^2 \theta \\ 0 & 0 & 0 & a_\phi \sin (2 \theta) & -a_\phi \sin (2 \theta) \\0 & - a_\phi' \sin^2 \theta & -a_\phi \sin (2 \theta) & 0 & 0 \\ 0 & -a_\phi' \cos^2 \theta & a_\phi \sin (2 \theta) & 0 & 0 \end{pmatrix} \eqend{.} \label{fullF}
\end{equation}
The eigenvalues and eigenvectors of $F^\mu{}_{\nu}$ are given by
\begin{enumerate}
\item
 Eigenvalue 0, eigenvector $k_{(0)}^\mu=\frac{2}{\alpha} (a_\phi',0,0, -a_0', -a_0')$.
\item
Eigenvalues  $\pm \alpha$, where
\begin{equation}
  \label{eigvalalpha}
 \alpha= 2 \sqrt{-f a_\phi'^2+ \frac{\N}{\m} a'^2}\eqend{,}
\end{equation}
and $a'=a_0'+\om  a_\phi' $,
eigenvectors
\begin{equation}
\label{eigenvectk}
 k_{(\pm)}^\mu = \sqrt{\frac{2}{ f \m \N}}  \alpha  \left( - \N a', \mp f \N \alpha , 0, f \m a_\phi' - \N \omega a', f \m a_\phi' - \N \omega a' \right)\eqend{.}
\end{equation}
\item
Eigenvalues $\pm \mathi \beta$,
where $\beta =  \dfrac{2 f a_\phi}{\m}$
with eigenvectors
\begin{equation}
\ell_{(\pm)}^\mu = \sqrt{f/  {2 \m} } \left( 0, 0, \pm \mathi, - \cot \theta, \tan \theta \right) \eqend{.}
\end{equation}
\end{enumerate}
The above eigenvectors can be used to construct a null basis
\begin{equation}
\begin{gathered}
\bl = \bm_{(0)} = \bk^{(+)} \eqend{,} \qquad \bn = \bm_{(1)} = - \bk^{(-)} \eqend{,} \qquad \bm_{(2)} = \bk^{(0)} \eqend{,}  \label{eqframe1}
\\
\bm_{(3)} = \sqrt{2} \, \Re \bl^{(+)} \eqend{,} \qquad \bm_{(4)} = \sqrt{2} \, \Im \bl^{(+)} \eqend{.}
\end{gathered}
\end{equation}
The only non-vanishing scalar products are then $\bl \cdot \bn = 1$ and $\bm_{(i)} \cdot \bm_{(j)} = \delta_{ij}$ for $i,j = 2,3,4$ and the metric reads
\begin{equation}
	g_{\mu\nu} = 2 \ell_{(\mu} n_{\nu)} + \delta_{ij} m^{(i)}_\mu m^{(j)}_\nu \eqend{.}
\end{equation}
In this basis, the Maxwell tensor reads
\begin{equation}
F_{\mu\nu} = 2 \alpha \ell_{[\mu} n_{\nu]} - 2 \beta m^{(3)}_{[\mu} m^{(4)}_{\nu]} \eqend{,}
\label{DtensorF}
\end{equation}
and the stress-energy tensor has the form
\begin{equation}
\begin{split}
T_{\mu\nu} &= F_{\mu\rho} F_\nu{}^\rho - \frac{1}{4} g_{\mu\nu} F_{\rho\sigma} F^{\rho\sigma} \\
&= \frac{1}{2} \left( \alpha^2 - \beta^2 \right) m^{(2)}_\mu m^{(2)}_\nu + \left( \alpha^2 + \beta^2 \right) \left( m^{(3)}_{\mu} m^{(3)}_{\nu} + m^{(4)}_{\mu} m^{(4)}_{\nu} -2 \ell_{(\mu} n_{\nu)} \right) \eqend{.} \label{DtensorT}
\end{split}
\end{equation}

Since both tensors $\bF$ and $\bT$ in \eqref{DtensorF} and \eqref{DtensorT} contain only boost weight zero terms, they are both algebraically special, more precisely of type~D in the algebraic classification of tensors \cite{Milsonetal05}. Note that this property holds off-shell (\emph{i.e.}, without imposing the field equations).

The Einstein equations (with $4 \pi G_5 = 1$)
\begin{equation}
R_{\mu\nu} - \frac{1}{2} R g_{\mu\nu} = 2 T_{\mu\nu} \eqend{,} \qquad R_{\mu\nu} = 2 T_{\mu\nu} - \frac{2}{3} g^{\rho\sigma} T_{\rho\sigma} g_{\mu\nu} \eqend{,}
\end{equation}
take the form
\begin{equation}
\begin{split}
R_{\mu\nu} &= - \frac{4}{3} \left( 2 \alpha^2 + \beta^2 \right) \ell_{(\mu} n_{\nu)} + \frac{2}{3} \left( \alpha^2 - \beta^2 \right) m^{(2)}_\mu m^{(2)}_\nu \\
&\quad+ \frac{2}{3} \left( \alpha^2 + 2 \beta^2 \right) m^{(3)}_\mu m^{(3)}_\nu + \frac{2}{3} \left( \alpha^2 + 2 \beta^2 \right) m^{(4)}_\mu m^{(4)}_\nu \eqend{.} \label{eqRicci}
\end{split}
\end{equation}
The Ricci tensor is thus of type~D on-shell (\emph{i.e.}, assuming that the Einstein equations hold).

Consider a general vector $\bv = v_0 \bn + v_1 \bl + v_2 \bm_{(2)} + v_3 \bm_{(3)} + v_4 \bm_{(4)}$ in the above null basis. We want to identify all null vectors  $\bv$ obeying $R_{ab} v^a v^b=0.$ Provided $\alpha \ne 0$, from the null condition and Equation~\eqref{eqRicci} it follows
that $v_2=v_3=v_4=0$ and $v_0 v_1=0$. Up to a factor, $\bl$ and $\bn$ are thus the only null vectors obeying $R_{ab} v^a v^b=0$.
The exceptional case $\alpha = 0$, after using the conservation
laws~\eqref{eq:reg-eom-cons}, corresponds to
\begin{equation}
	-f a_\phi'^2 + \frac{\hat{Q}^2 f^2}{\N \m} = 0 .
\end{equation}
This condition looks independent from the reduction of the equations of
motion to the constrained BVP~\eqref{eq:reg-eom-bvp}
and~\eqref{eq:reg-eom-constraint}, so it is likely inconsistent with
them. Formally, this can be checked by substituting the leading order
asymptotic expansion expansion~\eqref{eq:reg-sol-inf}. Setting to zero
the resulting leading $\bigo{\R^{-3}}$ coefficient gives $\hat{Q}=0$,
meaning that only the uncharged case allows $\alpha=0$.

The null congruence $\bl$ (and $\bn$) is geodesic if and only if
\begin{equation}
\label{geodcond}
a_\phi' a_0'' - a_0'a_\phi'' = 0 \eqend{.}
\end{equation}
This condition also looks independent from the reduction of the equations of
motion to the constrained BVP~\eqref{eq:reg-eom-bvp}
and~\eqref{eq:reg-eom-constraint}, so it is likely inconsistent with them.
Formally, this can be checked by continuing the
expansion~\eqref{eq:reg-sol-inf} by two more subleading orders
(as we have done in~\eqref{eq:reg-sol-inf4})
and plugging it into the desired
geodesic condition~\eqref{geodcond}. Setting to zero any
non-vanishing terms in the resulting asymptotic series then imposes
non-trivial constraints on the parameters which determine the general
solution (namely $\hat{M}$, $a_\phi^{(-1)}$ and $\r_\pm, \hat{J},
\hat{Q}$, as discussed in Section~\ref{sec:af}). The first non-trivial
term in the expansion of~\eqref{geodcond} appears at order
$\bigo{\R^{-6}}$. Setting to zero the coefficients of $\R^{-6}$ and
$\R^{-7}$ determines the $\hat{M}$ and $a_\phi^{(-1)}$ parameters,
while plugging these values into the coefficient of $\R^{-8}$ and
setting it to zero results in the simple constraint
\begin{equation}
	\hat{J}^3 \hat{Q}^5 = 0 \eqend{.} \label{geodcond2}
\end{equation}
In other words, the geodesic condition~\eqref{geodcond} on $\bl$ cannot
hold whenever both $\hat{J}\ne 0$ and $\hat{Q}\ne 0$.

This result, together with Proposition~\ref{prop_GKS_geod}, clearly explains why the metric for a charged rotating black hole in five dimensions cannot be written in the Kerr--Schild form \eqref{GKS} with a geodesic $\bk$, while the limiting cases with either $\hat{Q}=0$ or $\hat{J}=0$ are compatible with this form.

\subsubsection{Electromagnetic test field}
\label{sec_testfield}
It is of also of interest to compare the results of the previous section with the cases of a test field and the weak-field limit.
It is well known that in a vacuum spacetime, Killing vectors can be used to construct electromagnetic test fields obeying vacuum Maxwell equations in the background spacetime  \cite{Wald74}. The contribution of the test field to the energy-momentum tensor is neglected and thus test fields have a physical meaning only in the weak-field limit.
Electromagnetic test fields for the five-dimensional  Myers--Perry black hole were analyzed in \cite{AliFro04}. Additional properties of test fields in various five-dimensional black hole/ring spacetimes were studied in \cite{OrtPra06}. In particular, it has been shown in \cite{OrtPra06} that for the  Myers--Perry black hole, null eigenvectors of the test field are aligned with the WANDs of the background vacuum  Myers--Perry metric. The Weyl tensor and the Maxwell tensor are in this case both of type D and aligned.

The Maxwell tensor corresponding to the test field potential $A^\mu = \hat \alpha \delta^{\mu}_t$ reads
\begin{equation}
\label{testF}
\resizebox{0.93\hsize}{!}{$
F_{\mu\nu} = \dfrac{\hat \alpha \hat M}{(\R+a^2)^2}
\begin{pmatrix} 0 & 1 & 0 & 0 & 0 \\
-1 & 0 & 0 & a \sin^2 \theta & a \cos^2 \theta \\
0 & 0 & 0 & -a (\R+a^2) \sin(2 \theta)  & a (\R+a^2) \sin(2 \theta) \\
0 & -a \sin^2 \theta & a (\R+a^2) \sin(2 \theta) & 0 & 0\\
0 & -a \cos^2 \theta & -a (\R+a^2) \sin(2 \theta) & 0 \end{pmatrix} \eqend{,}
$}
\end{equation}
and the eigenvalues and corresponding null eigenvectors of $F^\mu{}_\nu$ are given by
\begin{equation}
\label{eigvalalphatest}
\pm \alpha_\text{T} = \pm \frac{2 \hat \alpha \hat M \sqrt{\R} }{(\R+a^2)^2} \eqend{,}
\end{equation}
\begin{equation}
\label{eigenvectktest}
k_{\text{T}(\pm)}^\mu = \left(a+\frac{\R}{a} , \mp \frac{2 \sqrt{\R}}{a (a^2+\R)} \left( a^4+ 2 a^2 \R + \R (\R- \hat M) \right), 0,1,1 \right) \eqend{,}
\end{equation}
where the subscript T stands for ``test field''.

\subsubsection{Behaviour for small charge and comparison with the test field}

Let us introduce the function $b$ by
\begin{equation}
a_\phi = \hat Q b \eqend{.}
\end{equation}
Then the field equations \eqref{eq:reg-eom-bvp} and constraint \eqref{eq:reg-eom-constraint}
contain only even powers of $\hat Q$, and for small charge we can expand the metric and potential functions in terms of $\hat{Q}^2$:
\begin{equation}
f = \sum_{k=0}^\infty f_{(k)} \frac{\hat{Q}^{2k}}{k!} \eqend{,} \quad \m = \sum_{k=0}^\infty \m_{(k)} \frac{\hat{Q}^{2k}}{k!} \eqend{,} \quad b = \sum_{k=0}^\infty b_{(k)} \frac{\hat{Q}^{2k}}{k!} \eqend{.}
\end{equation}
By neglecting second and higher powers of $\hat{Q}$, the field equations
\eqref{eq:reg-eom-bvp} and constraint \eqref{eq:reg-eom-constraint} for the
leading terms $f_{(0)}$, $\m_{(0)}$ and $b_{(0)}$ reduce to the
vacuum field equations for the uncharged Myers--Perry black hole, plus one
additional equation for $b_{(0)}$:
\begin{equation}
\label{eqb0}
\m_{(0)} (b_{(0)}' \m_{(0)})' - b_{(0)} f_{(0)} = 2 \hat{J} \frac{f_{(0)} \m_{(0)}^2}{\N^2}  \eqend{.}
\end{equation}
The Functions $f_{(0)}$ and $\m_{(0)}$ thus correspond to the vacuum Myers--Perry metric~\eqref{MPmetric}, and Equation~\eqref{eqb0} is solved by
\begin{equation}
b_{(0)} = \frac{\hat{J}}{\hat{M} (a^2 + R)} \eqend{.}
\end{equation}
Setting $\hat \alpha= \hat Q/ \hat M$, the eigenvalue $\alpha$ \eqref{eigvalalpha} of the Maxwell tensor reduces to the eigenvalue $\alpha_\text{T}$~\eqref{eigvalalphatest} of the test field and the eigenvectors~\eqref{eigenvectk} reduce to the test field eigenvectors~\eqref{eigenvectktest}.

Therefore, in the weak field limit the Maxwell tensor \eqref{fullF} reduces to the test field tensor~\eqref{testF} and is aligned with the (vacuum) background Weyl tensor. As we will see in the next section, this alignment of the Weyl and electromagnetic tensors does not survive beyond the weak field approximation.

\subsection{Algebraic type of the Weyl tensor}

\subsubsection{Weyl type of the metric ansatz}

It is well-known that the Kerr--Newman black hole in four dimensions is an algebraically special solution of Weyl type D.
As we will point out below, for the five-dimensional charged rotating black hole this does not hold anymore.
The algebraic classification of the Weyl tensor in higher dimensions is based on the alignment of a null vector, the so-called Weyl aligned null direction (WAND),
where the definition of the Weyl types is linked with the multiplicity of such a WAND (see for example~\cite{OrtPraPra12rev} for a review).
The standard approach to determine the algebraic type of a given Weyl tensor is to find an aligned null frame in which the vanishing components of the Weyl tensor then indicate its type.
Here, we employ the higher-dimensional extension of the Bel--Debever criteria \cite{Ortaggio:2009sb}, another simpler and frame-independent method to answer the question whether the five-dimensional charged rotating black hole solution is algebraically special.
The Bel--Debever criterion for a specific algebraic type is a polynomial equation involving the Weyl tensor and an unknown null vector $\bl$, which is the respective WAND of given multiplicity.
For instance, the Weyl tensor is of algebraic type II if and only if there exists a null vector $\bl$ satisfying
\begin{equation}\label{eq:BelDebeverII}
  \mathcal{C}^\text{II}_{\mu\nu\rho\sigma\tau} \equiv \ell_{[\sigma} C_{\mu]\lambda[\nu\rho} \ell_{\tau]} \ell^\lambda = 0 \eqend{.}
\end{equation}
Taking the regular metric ansatz \eqref{eq:reg-ansatz} in terms of the radial coordinate $\R$ into account, it is convenient to start with a general $\bl$ of the form
\begin{equation}
  \ell^\mu = \left(-\frac{\hat\ell^t}{f}, 2 \sqrt{\frac{\N}{\m}} \hat\ell^\R, \frac{\hat\ell^\theta}{\sqrt{\m}}, \sqrt{\frac{\m}{f\N}} \hat\ell^\phi - \frac{\hat\ell^t \varpi}{f}, \sqrt{\frac{\m}{f\N}} \hat\ell^\psi - \frac{\hat\ell^t \varpi}{f}\right) \eqend{,}
\end{equation}
which simplifies the condition for $\bl$ being null:
\begin{equation}
\begin{split}
&8 \left[ (\hat\ell^t)^2 - (\hat\ell^\R)^2 - (\hat\ell^\theta)^2 \right] - \left[ 3 (\hat\ell^\phi)^2 + 2 \hat\ell^\phi \hat\ell^\psi + 3 (\hat\ell^\psi)^2 \right] \\
&\quad + 4 \left[ (\hat\ell^\phi)^2 - (\hat\ell^\psi)^2 \right] \cos(2\theta) - (\hat\ell^\phi - \hat\ell^\psi)^2 \left[ \cos (4\theta) + \frac{2\m^2}{f\N} \sin^2 (2\theta) \right] = 0 \eqend{.}
\end{split}
\end{equation}
The Bel--Debever criterion \eqref{eq:BelDebeverII} for the Weyl tensor of our ansatz \eqref{eq:reg-ansatz} is considerably complicated.
It turns out that the simplest component of the rank-5 tensor $\mathcal{C}^\text{II}_{\R t\phi\theta\psi}$ factorizes and yields the condition
\begin{equation}
  \left[(\hat\ell^\theta)^2 + (\hat\ell^\R)^2\right] \left[\hat\ell^t \left( \N f' + f \N' - 2 \frac{f \N}{\m} \m' \right) + \sqrt{\frac{f \N^3}{\m}} (\hat\ell^\phi \sin^2\theta + \hat\ell^\psi \cos^2\theta) \varpi' \right]=0 \eqend{,}
\end{equation}
where a prime denotes a derivative with respect to $\R$. We thus obtain two branches: either the first bracket vanishes, which entails $\ell^\theta = \ell^\R = 0$, or the second bracket vanishes.
Both conditions ensure that the component $\mathcal{C}^\text{II}_{\R t\phi\theta\psi}$ vanishes, and for each branch we can simplify the criterion \eqref{eq:BelDebeverII} by substituting back the corresponding condition.
We then repeat this procedure: picking up the remaining simplest component which factorizes, and plugging the condition in each new subbranch into the criterion \eqref{eq:BelDebeverII},
until all the components of the criterion are satisfied.
In this way, we obtain at the end 109 sets of conditions on the metric functions and the components of the WAND $\bl$ under which the Bel--Debever criterion for type II \eqref{eq:BelDebeverII} is satisfied.
However, all these sets are contained in the following 9 disjunct cases:
\begin{subequations}\label{eq:typeIIconds}
\begin{align}
\begin{split}
  1:\quad & (\hat\ell^t)^2 = (\hat\ell^\R)^2 + (\hat\ell^\theta)^2 + (\hat\ell^\phi)^2 \sin^2\theta + (\hat\ell^\psi)^2 \cos^2\theta \eqend{,} \\
    & f\N = \m^2, \quad \varpi' = 0 \eqend{,}
    \quad f'' = \frac{(f')^2}{4f} + f \frac{4 \N (\N'' + 1) - 3 (\N')^2}{12 \N^2} \eqend{,}
\end{split} \\
\begin{split}
  2:\quad & \hat\ell^\theta = 0 \eqend{,} \quad \hat\ell^\psi = \hat\ell^\phi \eqend{,} \quad (\hat\ell^t)^2 = (\hat\ell^\R)^2 + (\hat\ell^\phi)^2 \eqend{,} \\
    & \left(\frac{f\N}{\m^2}\right)' = \varpi' = 0 \eqend{,}
    \quad f'' = \frac{f^2}{3 \m^2} + \frac{(f')^2}{4 f} + f \frac{4 \N \N'' - 3 (\N')^2}{12 \N^2} \eqend{,}
\end{split} \\
\begin{split}
  3:\quad & \hat\ell^\theta = 0 \eqend{,} \quad \hat\ell^\phi = 0 \eqend{,} \quad \hat\ell^\psi = 0 \eqend{,} \quad (\hat\ell^t)^2 = (\hat\ell^\R)^2 \eqend{,} \\
    & \left(\frac{f\N}{\m^2}\right)' = \varpi' = 0 \eqend{,}
    \quad f'' = \frac{5 f^2}{3 \m^2} + \frac{(f')^2}{4 f} + f \frac{4 \N (\N'' -4) - 3 (\N')^2}{12 \N^2} \eqend{,}
\end{split} \\
\begin{split}
  4:\quad & \hat\ell^\theta = 0 \eqend{,} \quad \hat\ell^\phi = 0 \eqend{,} \quad \hat\ell^\psi = 0 \eqend{,} \quad (\hat\ell^t)^2 = (\hat\ell^\R)^2 \eqend{,} \\
    & \left(\frac{f\N}{\m^2}\right)' = \varpi' = 0 \eqend{,}
    \quad f'' = - \frac{f^2}{3 \m^2} + \frac{(f')^2}{4 f} + f \frac{4 \N (\N'' + 2) - 3 (\N')^2}{12 \N^2} \eqend{,}
\end{split} \\
\begin{split}
  5:\quad & \hat\ell^\theta = 0 \eqend{,} \quad \hat\ell^\phi = 0 \eqend{,} \quad \hat\ell^\psi = 0 \eqend{,} \quad (\hat\ell^t)^2 = (\hat\ell^\R)^2 \eqend{,} \\
    & \left(\frac{f\N}{\m^2}\right)' = 0 \eqend{,} \quad \varpi'' = \frac{3 (\N f' - f \N') \varpi'}{4 f \N} \eqend{,}
\end{split} \\
\begin{split}
  6:\quad & \hat\ell^t = \hat\ell^\phi \sqrt{\frac{\N}{f\m}} \varpi \eqend{,} \quad \hat\ell^\theta = 0 \eqend{,} \quad \hat\ell^\psi = \hat\ell^\phi \eqend{,}
    \quad (\hat\ell^\R)^2 = (\hat\ell^\phi)^2 \left( \frac{\N}{f\m} \varpi^2 -1 \right) \eqend{,} \\
    & \left(\frac{f\N\varpi}{\m^2}\right)' = 0 \eqend{,}
    \quad 4 \m^2 f'' = f^2 + \frac{\m f' (2 \m f' - 3 \N \varpi \varpi')}{f} - f ( (\m')^2 - 2 \m \m'') + 3 \N \varpi  \m' \varpi' \\
    &\quad - \m (f' \m' - 2 \N \varpi \varpi'') \eqend{,}
\end{split} \\
\begin{split}
  7:\quad & \hat\ell^t = \hat\ell^\phi \sqrt{\frac{\N}{f\m}} \varpi \eqend{,} \quad \hat\ell^\theta = 0 \eqend{,} \quad \hat\ell^\psi = \hat\ell^\phi \eqend{,}
    \quad (\hat\ell^\R)^2 = (\hat\ell^\phi)^2 \left( \frac{\N}{f\m} \varpi^2 - 1 \right) \eqend{,} \\
    & \left(\frac{f\N\varpi}{\m^2}\right)' = 0 \eqend{,}
    \quad (\N \varpi^2 - f \m) \varpi'' = \frac{8 f^3 \varpi + 3 \m \N \varpi^2 f' \varpi'}{4 f \m} - f \m \frac{2 \varpi^2 -  \varpi \N' \varpi' + \N (\varpi')^2}{\N \varpi } \\
    &\quad - \varpi' \frac{6 \m f' + \varpi (3 \varpi \N' - 7 \N \varpi')}{4} \eqend{,}
    \quad 12 f'' = \frac{20 f^2}{\m^2} + \frac{3 (f')^2}{f} - 4 \varpi' \left(\frac{3 f'}{\varpi} - \frac{5 \N \varpi'}{\m}\right) \\
    &\quad - \frac{3 f (\N')^2}{\N^2} + 2 f \frac{5 \N' \varpi' - 2 \varpi (4 - \N'')}{\N \varpi} - f \frac{11 (\varpi')^2 - 12 \varpi \varpi''}{\varpi^2} \eqend{,}
\end{split} \\
\begin{split}
  8:\quad & \hat\ell^\R = 0 \eqend{,} \quad \hat\ell^\theta = 0 \eqend{,} \quad \hat\ell^\psi = \hat\ell^\phi \eqend{,} \quad (\hat\ell^t)^2 = (\hat\ell^\phi)^2 \eqend{,}
    \quad \varpi' = \frac{\hat\ell^\phi}{\hat\ell^t} \sqrt{\frac{f^3 \N}{\m^3}} \left(\frac{\m^2}{f \N}\right)' \eqend{,} \\
    & 6 f'' = \frac{f^2}{\m^2} + 3 \frac{(f')^2}{f} -  \frac{f' \N'}{N} + \frac{2 f (\N')^2 }{\N^2} - 3 f \frac{ (\m')^2 - 2 \m \m''}{\m^2}
    - f \frac{ \m' \N' + 2 \m \N''}{\m \N} \eqend{,}
\end{split} \\
\begin{split}
  9:\quad & \hat\ell^t = - \hat\ell^\phi \sqrt{\frac{f \N^3}{\m}} \frac{\varpi'}{\N f' + f \N' - 2 \frac{f\N}{\m} \m'} \eqend{,} \quad \hat\ell^\theta = 0 \eqend{,}
    \quad \hat\ell^\psi = \hat\ell^\phi \eqend{,} \quad (\hat\ell^\R)^2 + (\hat\ell^\phi)^2 = (\hat\ell^t)^2 \eqend{,} \\
    & \frac{f''}{2f} =  \frac{(\m')^2}{\m^2} - \frac{2 \m' \N'}{\m\N} + \frac{\m''}{\m}
    + \left(\frac{2f'}{f} + \frac{\varpi''}{\varpi'}\right) \left( \frac{f'}{2f} +  \frac{\N'}{2\N} - \frac{\m'}{\m} \right)
    +  \frac{(\N')^2}{\N^2} - \frac{\N''}{2\N} \eqend{,} \\
    &2 (2 f \N \m' - \m (\N f' + f \N')) \m'' =  \frac{2 \m^2 \N (f')^3 }{f^2} + \frac{2 f \N (\m')^3 }{\m} + f (\N f' - (\m')^2 \N') \\
    &\quad - f^2 \frac{2 \N \m' - \m \N'}{\m} - \m f' \frac{-2 \m f' \N' - 3 \N^2 (\varpi')^2 + \N( 5 f' \m' + 4 \m f'')}{f} - \m \N' (f' \m' + 4 \m f'') \\
    &\quad + \N \m' (f' \m' + 8 \m f'') - \N^2 \varpi' (3 \m' \varpi' + 2 \m \varpi'') \eqend{.}
\end{split}
\end{align}
\end{subequations}
In other words, the metric ansatz \eqref{eq:reg-ansatz} is of Weyl type~II if and only if the conditions of any of these 9 cases are met.
Nevertheless, it is not clear so far whether these conditions could be satisfied on-shell.
First, we recall that the uncharged and non-rotating limits of the charged rotating solution lead to known exact solutions,
namely the 5-dimensional Myers--Perry black hole and the 5-dimensional charged Schwarzschild--Tangherlini black hole, respectively,
 which are given in Section~\ref{sec:exact_solutions}.
Although both special solutions are of Weyl type~D, they satisfy the type~II conditions from \emph{distinct} cases which suggests
that the charged rotating solution is of more general algebraic type.
Specifically, the Myers--Perry black hole fulfills the conditions of case 9, whereas the charged Schwarzschild--Tangherlini black hole meets the conditions of case 5.

Now, let us discuss the compatibility of the conditions \eqref{eq:typeIIconds} with a general charged rotating solution of the Einstein--Maxwell equations.
The conditions of cases 1, 2, 3 and 4 are satisfied only in the non-rotating limit, since for $\varpi' = 0$
we obtain $\hat J = 0$ from the conservation law~\eqref{eq:reg-eom-cons-omega} together with the BVP~\eqref{eq:reg-eom-bvp-a}.
To analyze the remaining cases, we make use of the on-shell asymptotic expansions~\eqref{eq:reg-sol-inf4} of the functions $f$, $\m$ and $a_\phi$, where without loss of generality we set $\r_-^2=0$.
In the case 5, plugging the asymptotic expansions into $( \N f/\m^2 )'= 0$, the numerator given by
$\N f  = \R^2 - (\hat{M} + \rH^2)\R + \frac{1}{2} \hat{M}(\hat{M} + \rH^2) + \frac{4}{3}\hat Q^2 + \bigo{\R^{-1}}$
and the denominator given by
$\m^2  = \R^2 - (\hat{M} + \rH^2)\R + \frac{1}{4} (\hat{M} + \rH^2)^2 + \frac{8}{3} \hat Q^2 + \bigo{\R^{-1}}$
differ at order $\R^0$. Therefore, $\N f/\m^2$ in general depends on $\R$ and the conditions of case 5 are not fulfilled.
Analogously, in cases 6 and 7, from the first terms of the asymptotic expansion of $( \N f \varpi/\m^2 )' = 0$
it is obvious that the fraction of $\N f \varpi = \hat J + \bigo{\R^{-1}}$
and $\m^2 = \R^2 + \bigo{\R}$ is not independent of $\R$, such that also this condition is not met for a general solution.
Substituting the asymptotic expansion \eqref{eq:reg-sol-inf4} in the condition $\varpi' = \pm \sqrt{\N f^3/\m^3} [ \m^2/(\N f) ]'$ of case 8
we obtain $2 \hat J/\R^3 = \pm ( 16 \hat Q^2 - 3 \hat M^2 + 3 \rH^4 )/(6 \R^{7/2}) + \bigo{\R^{-4}}$, requiring $\hat J = 0$.
Lastly, in case 9 the leading order term in its condition reads $- 4 \hat J \R  + \bigo{\R^0} = 0$, and again requires that $\hat J = 0$, such that none of the conditions can be fulfilled for a general charged rotating solution.

\subsubsection{Weyl type on the horizon}

We have shown above that in the bulk the charged Myers--Perry metric is algebraically general.
In this section, we  determine the algebraic type of the metric on the horizon.
The form~\eqref{eq:reg-ansatz-kruskal} is manifestly regular at the
horizons and the bifurcation sphere, but it is more cumbersome than
necessary. In Kruskal-like coordinates, an a~priori regular ansatz may
be parameterized as follows, where any function with implicit coordinate
dependence depends only on the product $UV$ and have non-vanishing
leading terms of order zero, \emph{e.g.}, $F = F(UV) = F(0) + \bigo{UV}$:
\begin{multline}
\d{s}^2 = 2 F \d{U}\d{V}
+ N [W (U\d{V}-V\d{U}) - \sin^2\theta\d\phi - \cos^2\theta\d\psi]^2
\\
+ M \d\theta^2
+ M \sin^2\theta \cos^2\theta (\d\phi - \d\psi)^2 \eqend{,} \label{xKS}
\end{multline}
\begin{equation}
\resizebox{0.93\hsize}{!}{$
g_{\mu\nu} =
\begin{pmatrix}
V^2 N W^2 & F - UV N W^2 & 0 & V N W \sin^2\theta & V N W \cos^2\theta \\
F - UV N W^2 & U^2 N W^2 & 0 & -U N W \sin^2\theta & -U N W \cos^2\theta \\
0 & 0 & M & 0 & 0 \\
V N W \sin^2\theta & -U N W \sin^2\theta & 0 &
(N\sin^2\theta + M\cos^2\theta) \sin^2\theta & (N-M) \sin^2\theta\cos^2\theta \\
V N W \cos^2\theta & -U N W \cos^2\theta & 0 &
(N-M) \sin^2\theta\cos^2\theta & (N\cos^2\theta + M\sin^2\theta) \cos^2\theta \\
\end{pmatrix} \eqend{.}
$}
\end{equation}
To determine the type of the Weyl tensor on the horizon, we employ the null coframe $\bl=\d{U}$, $\bn=\d{V}$, $\bm^{(2)}=\d{\theta}$, $\bm^{(3)}=\d\phi - \d\psi$,  $\bm^{(4)}=W (V\d{U}-U\d{V})+\d\phi $ (up to normalization).

Calculating all boost-weight +2 components of the Weyl tensor $C^{\alpha \beta \gamma \delta} \ell_{\alpha} m^{(i)}{}_\beta \ell_{\gamma} m^{(j)}{}_\delta$, we find that they all vanish for $U=0$. Furthermore, all boost-weight +1 components vanish there as well. It follows that $\bl$ is a multiple WAND at $U=0$. Similarly, $\bn$ is a multiple WAND at $V=0$. Therefore, the spacetime is at least of Weyl type~II at either of the horizons ($V=0$ or $U=0$) and type~D at the bifurcation sphere ($U=V=0$). The same results also hold for the Riemann tensor.
Note that these observations hold even off-shell, without imposing the Einstein--Maxwell equations.

\section{Discussion} \label{sec:discuss}

We have studied various mathematical aspects of a charged rotating black hole with two equal-magnitude angular momenta in five dimensions.

In Section~\ref{sec:reg}, we proposed a metric ansatz \eqref{eq:reg-ansatz} which is regular on the horizon. Furthermore, for this metric ansatz the Einstein--Maxwell system~\eqref{eq:reg-eom-bvp} and~\eqref{eq:reg-eom-constraint} is autonomous, which is technically convenient.
We studied geometric regularity conditions for the Einstein--Maxwell system~\eqref{eq:reg-eom-bvp} following from asymptotic flatness with the absence of incoming and outgoing radiation, as well as regularity at the horizon. By comparing the solution at the singular end-points with these conditions, we showed that the boundary value problem is well-posed.

In Section~\ref{sec:algebraic}, we performed the algebraic classification of the curvature tensors and the electromagnetic tensor $\bF$. We found that the Weyl tensor is algebraically general in the bulk, type~II on the horizon, and type~D on the bifurcation sphere. The electromagnetic field strength tensor $\bF$ and the stress-energy tensor $\bT$ are both of type~D off-shell, while the Ricci tensor is type~D on-shell. These algebraic properties are inconsistent with the Kerr--Schild form~\eqref{GKS} of the metric with geodesic $\bk$, such that the charged rotating black hole with two equal-magnitude angular momenta in five dimensions cannot be described by such a Kerr--Schild metric. On the other hand, changing to coordinates appropriate for studying the on-horizon properties of the spacetime, the metric~\eqref{xKS} takes a form somewhat reminiscent of the Kerr--Schild form. In this case, however, the background metric is a four-dimensional type~D warped product metric, and the ``Kerr--Schild'' vector is not null.

\section*{Acknowledgments}

This work has been supported by the Research Plan RVO
67985840 and Czech Science
Foundation Grant No. 19-09659S (I.K., T.M., and V.P.). M.B.F. is supported by the Deutsche Forschungsgemeinschaft (DFG, German Research Foundation) --- project no. 396692871 within the Emmy Noether
grant CA1850/1-1 and project no. 406116891 within the Research Training Group RTG 2522/1.
In addition, I.K. was partially supported by the Praemium Academiae of M.~Markl.

\appendix

\section{Comparison with the work of Fan, Liang, Mei} \label{sec:flm}

In~\cite{fan-liang-mei}, Fan \emph{et al.}\ looked at the same metric
ansatz, introduced by Kunz \emph{et al.}~\cite{kunz-etal, nl-5d}, for a
5-dimensional charged rotating black hole that we are considering in
this work. They claim to have reduced the corresponding Einstein
equations to only two scalar variables (their $Z$ and $A_\phi$). In
contrast, in Section~\ref{sec:reg}, the most we could reduce our system
to is the three scalar variables $f$, $\m$, $a_\phi$, in
Equations~\eqref{eq:reg-eom-bvp} and~\eqref{eq:reg-eom-constraint}.
Unfortunately, a closer comparison reveals some inconsistencies in their
formulas.

In more detail, the parametrization of the ansatz used
in~\cite{fan-liang-mei} is
\begin{subequations}
\begin{multline}
	\total s^2 = - F \total t^2 + \frac{\total R^2}{W} + R^2 \left( \frac{\total x^2}{1-x^2} + (1-x^2) \total \phi^2 + x^2 \total \psi^2 \right) \\
	+ N \left[ (1-x^2) \total \phi + x^2 \total \psi \right]^2 - 2 B \left[ (1-x^2) \total \phi + x^2 \total \psi \right] \total t \eqend{,}
\end{multline}
\begin{equation}
A = A_t \total t + A_\phi \left[ (1-x^2) \total \phi + x^2 \total \psi \right] \eqend{.}
\end{equation}
\end{subequations}
where a comparison with our parametrization~\eqref{eq:reg-ansatz}
reveals the dictionary
\begin{subequations}
\begin{align}
	x &= \cos\theta \eqend{,} \\
	r &= \sqrt{\m/f} \eqend{,} \\
	F &= f - \frac{\N}{\m} \varpi^2 \eqend{,} \\
	W &= \N \left(\frac{\m'}{\m} - \frac{f'}{f}\right)^2 \eqend{,} \\
	N &= \frac{\N}{\m} - \frac{\m}{f} \eqend{,} \\
	B &= \frac{\N}{\m} \varpi \eqend{,} \\
	A_t &= a_0 \eqend{,} \\
	A_\phi &= a_\phi \eqend{,}
\end{align}
\end{subequations}
where all derivatives on the right-hand side are with respect to our
coordinate $\R = \r^2$. In the two equations,
\begin{align}
\tag*{Eq.~(3) of \cite{fan-liang-mei}}
	W &= \frac{16 Z (c_0+Z)}{r^2 (\del_r Z)^2} \eqend{,} \\
\tag*{Eq.~(4) of \cite{fan-liang-mei}}
	N &= \frac{Z}{r^2 F} - \frac{B^2}{F} - r^2 \eqend{,}
\end{align}
the first one introduces a new variable $Z=Z(r)$ and a constant $c_0$,
while the second one follows from this definition in conjunction with
the Einstein equations. Using our dictionary and simple algebraic
manipulation, we find that $Z$ and $c_0$ are uniquely determined by the
identities
\begin{equation}
	Z = \N = (\R-\r_+^2)(\R-\r_-^2) \eqend{,}
	\quad
	c_0 + Z = (\N'/2)^2 = \left(\R-\frac{\r_+^2+\r_-^2}{2}\right)^2 \eqend{,}
\end{equation}
where we have used the form of $\N$~\eqref{eq:reg-eom-cons} imposed by
the Einstein equations. Next, after introducing the variable
$\tilde{A}_\phi$ in their Equation~(7), their formula~(6) for $F$ is
equivalent to a combination of our~\eqref{eq:reg-eom-cons-omega}
and~\eqref{eq:reg-eom-bvp-a}. However, their final complicated
formula~(8) for $B$ turns out to be inconsistent. We have verified the
inconsistency by plugging in our general asymptotic
solution~\eqref{eq:reg-sol-inf4}, for any non-trivial values of the free
parameters $\hat{M}$ and $a_\phi^{(-1)}$. Unfortunately, the claimed
reduction by~\cite{fan-liang-mei} of the Einstein equations to just the
two variables $Z$ and $A_\phi$ cannot be correct without the validity of
their Equation~(8), which we have also verified cannot be saved by
undoing possible sign errors or other minor typos. Finally, the
asymptotic solution at infinity~(9) for $Z$ and $A_\phi$
in~\cite{fan-liang-mei} cannot be matched up to our asymptotic
solution~\eqref{eq:reg-sol-inf4} by any correspondence between their
free parameters ($m,a,s$) and ours ($\hat{M}, a_\phi^{(-1)}$), with an
inconsistency appearing already at the next subleading order.

\bibliographystyle{utphys-alpha}
\bibliography{chargedrot5d,chargedrot5dVojta}

\end{document}